\documentclass[preprint2,times,tighten]{aastex6}
\usepackage{txfonts}
\usepackage{graphicx}
\usepackage{xcolor}

\slugcomment{to appear in special issue ApJSS., 2018}

\shorttitle{Diffuse X-ray emission in the Cygnus OB2 association}
\shortauthors{Albacete Colombo et al.}

\begin{document}


\title{Diffuse X-ray emission in the Cygnus OB2 association.}

\author{J. F. Albacete Colombo\altaffilmark{1},
J. J. Drake\altaffilmark{2},
E. Flaccomio\altaffilmark{3},
N. J. Wright\altaffilmark{4},
V. Kashyap\altaffilmark{2},
M. G. Guarcello\altaffilmark{3},
K. Briggs\altaffilmark{5},
J. E. Drew\altaffilmark{6},
D. M. Fenech\altaffilmark{7},
G. Micela\altaffilmark{3},
M. McCollough\altaffilmark{2},
R. K. Prinja\altaffilmark{7},
N. Schneider\altaffilmark{8},
S. Sciortino\altaffilmark{3},
J. S. Vink\altaffilmark{9}
}
\email{e-mail: albacete.facundo@conicet.gov.ar}
\altaffiltext{1}{Universidad de R\'io Negro, Sede Atl\'antica, Viedma CP8500, Argentina.}
\altaffiltext{2}{Smithsonian Astrophysical Observatory, 60 Garden St., Cambridge, MA 02138, U.S.A}
\altaffiltext{3}{INAF-Osservatorio Astronomico di Palermo, Piazza del Parlamento 1, 90134 Palermo, Italy.}
\altaffiltext{4}{Astrophysics Group, Keele University, Keele, Staffordshire ST5 5BG, UK.}
\altaffiltext{5}{Hamburger Sternwarte, University of Hamburg, Gojenbergsweg 112, 21029, Hamburg, Germany.}
\altaffiltext{6}{School of Physics, Astronomy \& Mathematics, University of Hertfordshire, College Lane, Hatfield, Hertfordshire, AL10 9AB, UK.}
\altaffiltext{7}{Department of Physics and Astronomy, University College London, Gower Street, London WC1E 6BT, UK.}
\altaffiltext{8}{I. Physik. Institut, University of Cologne, D-50937 Cologne, Germany.}
\altaffiltext{9}{Armagh Observatory and Planetarium, College Hill, Armagh BT61 9DG, Northern Ireland, UK.}

\begin{abstract}
We present a large-scale study of diffuse X-ray emission in the nearby massive 
stellar association Cygnus OB2 as part of the Chandra Cygnus OB2 Legacy Program. We used 40 Chandra X-ray ACIS-I observations covering $\sim$1.0 deg$^2$. After removing 7924 point-like sources detected in our survey, background-corrected X-ray emission, the adaptive smoothing reveals large-scale diffuse X-ray emission. Diffuse emission was detected in the sub-bands Soft [0.5\,:\,1.2] and Medium [1.2\,:\,2.5], and marginally in the Hard [2.5\,:\,7.0] keV band. From X-ray spectral analysis of stacked spectra we compute a total [0.5-7.0 keV] diffuse X-ray luminosity of L$_{\rm x}^{\rm diff}\approx$4.2$\times$10$^{\rm 34}$ erg\,s$^{-1}$, characterized with plasma temperature components at kT$\approx$ 0.11, 0.40 and 1.18 keV, respectively.
The H{\sc i} absorption column density corresponding to these temperatures 
has a distribution consistent with N$_{\rm H}$ =\,0.43, 0.80 and 1.39 $\times$10$^{22}$ cm$^{-2}$. 
The extended medium band energy emission likely arises from O-type stellar winds thermalized by wind-wind collisions in the most populated regions of the association, while the soft band emission probably arises from less energetic termination shocks against the surrounding Interstellar-Medium.  Super-soft and Soft diffuse emission appears more  widely dispersed and intense 
than  the medium band emission. The diffuse X-ray emission is generally spatially coincident with  low-extinction regions that we attribute to the  ubiquitous influence of powerful stellar winds from massive stars and their interaction with the local Interstellar-Medium. Diffuse X-ray emission is volume-filling, rather than edge-brightened,  oppositely to other star-forming regions. We reveal the first observational evidence of X-ray haloes around some evolved massive stars.
\end{abstract}


\section{Introduction}
Since the early X-ray observations by {\it Einstein} and later {\it ROSAT}, and up to the last decade, 
the study of diffuse X-ray emission associated with stellar clusters and star-forming regions 
has presented considerable challenges. 
While in supernova remnants or large-scale massive structures, such as galaxy clusters, detections 
can be quite clear, confirmation of  diffuse X-ray emission from Galactic star-forming regions 
(SFRs) has often remained elusive. Several attempts were made to detect diffuse 
X-rays from young massive SFRs with ROSAT  \citep[e.g.,][]{Strickland1998}. However, the 
lack of adequate sensitivity and spatial resolution rendered it difficult to  
distinguish between genuine diffuse emission and X-ray emission from unresolved stars.
Limited by the available instrumentation, this subfield of astrophysics remained without major progress 
for more than 20 years.  This ended in 1999 with the launch of the {\it Chandra} X-ray Observatory  which
combines high sensitivity in the 0.5--8 keV energy range with spectacular spatial 
resolution ($\approx$ 0.5" on axis).  This combination greatly improved the  capacity to detect 
faint X-ray sources and  disentangle point source and true diffuse emission contributions 
in nearby Galactic SFRs. The first genuine discovery of diffuse X-ray emission
in a massive star forming region came from the Rosette and Omega 
Nebulae \citep{Townsley2003, Muno2006}.  Subsequently, \textit{XMM-Newton} observations of 
the Orion Nebula revealed it to be filled by soft X-ray-emitting (2-MK) plasma \citep{G2008}. 
More recently, in the context of the 
"Chandra Carina Complex Project" (CCCP) \citet{Townsley2011a} has published 
a milestone work for study of X-ray diffuse emission in Carina.

From a theoretical point of view, diffuse X-ray emission 
is expected to occur in some young stellar associations and SFRs due to supersonic 
stellar winds from massive stars that can produce dissipative shock waves in the 
local ISM, modifying astrophysical conditions of the molecular cloud material 
in the region. These shocks have been interpreted as evidence of non-radiative 
heating processes \citep{Polcaro1991} that can give rise to a number of interesting, 
though poorly explored, high-energy phenomena. However, the processes responsible for 
the production of X-ray diffuse emission are still not well understood,  especially where 
both thermal and non-thermal (NT) processes may be present.  
A usual key indicator of thermal diffuse X-ray emission is the presence 
of spectral lines \citep{Smith2001}, although NT interactions can also produce intense and relatively 
soft X-ray emission lines ($<$ 2keV) via  the Change Exchange (CXE) mechanism. 
Otherwise, a featureless spectrum without lines can originate either  from 
thermal processes (e.g., hot bremsstrahlung) or non-thermal electrons
via synchrotron emission or inverse Compton (IC) scattering processes.

The first detailed theoretical study was performed in a seminal paper by \cite{Weaver1977}.  
Further development of  models for X-ray emission from wind-blown bubbles was carried out by \cite{Canto2000} 
and \cite{Stevens2003}: they interpreted diffuse X-ray emission associated with massive clusters in 
terms of a collective cluster wind model (CWM). Winds from individual massive stars with mass loss 
rates of $\sim 10^{-6}$ M$_{\odot}$~yr$^{-1}$ and terminal speeds of 1600--2500 km~s$^{-1}$ collide, 
thermalize and expand supersonically into the  local ISM. This interaction acts as a 
precursor of an ``interstellar bubble'' that expands to a few tenths of a parsec around stars more 
massive than $\sim$8~M$_{\odot}$.  An approximate estimate for a typical single O and/or early B star 
luminosity and wind kinetic energy is $L_{{\rm bol}} \sim 10^{38\,-\,39}$ erg s$^{-1}$ and 
$L_{{\rm w}} \sim 10^{36\,-\,37}$ ergs s$^{-1}$. respectively.  The adopted typical efficiency of wind 
momentum to radiative conversion $\eta$ is $10^{-4}$ for interstellar shocks \citep{G2008} resulting in 
an X-ray diffuse emission luminosity $L_x^{{\rm diffuse}} \sim 10^{33\,-\,34}$ ergs s$^{-1}$.  

However models of such wind-blown bubbles predict much larger sizes than those reported by 
\cite{Bruhweiler2010} for the example of the Rosette Nebula. The discrepancy implies that the shocks 
involved should occur in the radiative (energy loss) regime. Otherwise the high wind speeds should 
produce strong $\sim$keV X-ray diffuse emission within regions of a few  parsec scale. At the same time, 
evidence for very large "superbubbles'' enclosing entire OB associations and shaped by multiple 
supernovae has been presented.  The particular case of the Cygnus superbubble, on a scale well 
in excess of 100~pc \citep{Cash1980}, is relevant here.  But \cite{Uyaniker2001} have argued that the 
extensive X-ray emission is better viewed as a superposition of a succession of separate regions at 
different distances. Accordingly, in Cygnus, an investigation of improving high-energy data can make 
a valuable contribution to clarifying our understanding, and we can access spatial scales of up to 
tenths of parsec.  

In this work, we identify truly diffuse X-ray emission in the 
Cygnus OB2 association, one of the most massive groups of young stars known in the  Galaxy.  
Assessments of its stellar complement have found that Cygnus OB2 contains well over 100 OB
stars \citep[e.g.,][]{Schulte1956, Massey1991, Comeron2002, Hanson2003, Wright2015a} 
and tens of thousands of lower-mass, pre-main sequence stars \citep[e.g.,][]{Albacete2007, 
Drew2008, Vink2008, Wright2009}. 
Cygnus OB2 lies at the center of the Cygnus\,X giant molecular cloud and star forming complex 
\citep{Schneider2006, Reipurth2008}, and is a source of feedback for the region \citep{Wright2012, Guarcello2016}. 
Its size and proximity make Cygnus\,OB2 an ideal environment to search for diffuse emission 
resulting from feedback into the environment from massive stars.

This study forms a part of the science exploitation of 
the {\em Chandra} Cygnus OB2 Legacy Survey (Drake et. al., this issue).   
This 1.1~Ms survey comprises a mosaic of {\em Chandra} ACIS-I pointings covering the central 
square degree of the association, which contains the majority of the massive young stellar association. 
The observations and the source
catalog are described in \citet{Wright2014}. Guarcello et al. (2017) have matched X-ray sources 
with available optical and infrared (IR) photometric catalogs, while Kashyap et al. (2017) have applied 
statistical methods to separate out approximately $\sim$6000 sources identified as association members 
from an additional $\sim 2000$ sources assessed as foreground and background interlopers.  Flaccomio et. al. (2018)
characterize the X-ray spectral properties of these populations, and discuss the results in the context of different 
emission models.
This thorough assessment of the point source content, combined with the deep and wide nature of the 
{\it Chandra} survey itself, provides a valuable dataset for a thorough exploration of X-rays from diffuse gas in the region.

In this article we focus on the analysis, detection and morphological 
description of diffuse X-ray emission in Cygnus\,OB2.  We discuss its origin and derive approximate 
astrophysical properties of the diffuse X-ray structures at large and small scales.\\
 
\section{X-ray observations and diffuse emission analysis}
\label{analysis_x} 

The study of the diffuse X-ray emission is of interest for assessing the total X-ray 
output of a region like Cygnus\,OB2, and will be important for understanding the X-ray 
emission from much more distant, unresolved stellar clusters.
The expected diffuse X-ray emission from Cyg~OB2 will be spread over a large
angular sky area ($\sim 1$ sq. deg.) and will have a  correspondingly  low 
surface brightness.  
An interesting rough comparison is the expected diffuse signal compared 
with the combined signal from detected point sources.
We can estimate the X-ray luminosity of the detected Cygnus\,OB2
stellar population by assuming that: $i)$ we detect all members with mass
$>1M_\odot$ \citep{Wright2015} $ii)$ the IMF of Cygnus OB2 is similar to that of the 
Orion Nebular Cluster (ONC), and $iii)$ Cygnus\,OB2 stars have the same X-ray activity as derived for
ONC stars by \cite{Preibisch2005}.  Adopting the ``lightly absorbed
optical sample'' as representative of the low-mass ONC population, 
we estimate that the Cygnus\,OB2 population of stars with mass between
1.0 and $10\,M_\odot$ is $\sim$10 times larger than the ONC one. We can then
scale the total luminosity of the ONC sample by this factor,
obtaining L$_x^{\rm LMS\,\,total}\sim 2\times 10^{34}$ erg\,s$^{-1}$ for the whole association,
which is comparable to what we expect for the diffuse emission. 
To deal successfully with the difficult task of extracting the diffuse emission signal, we made use of the specific data
analysis software ACIS-Extract (AE) \citep{Broos2012}, which is able to
remove the point source X-ray emission via the construction of 
``swiss-cheese" ACIS-I images with holes at the positions of detected point sources.
It computes the local point spread function (PSF) at each source position and masks its local contribution to
the observation (see section\,\ref{analysis_x}). However, the removal of
all detected point sources does not of course guarantee that the remaining X-ray
emission will be truly diffuse, since the summed
contribution of undetected 
faint sources could masquerade as diffuse X-ray emission. To assess how big this
problem is, we adopt the completeness limit of the survey in the central 0.5 sq.deg as
computed by \cite{Wright2015}. Assuming
that the distribution of X-ray luminosities at a given mass is the same
as in the ONC, as sampled by the  Chandra Orion Ultradeep Project (COUP) 
"lightly absorbed optical sample'', and that the population is 10 times as large (see above), we
estimate that the X-ray luminosity of the undetected population of
sources is $\sim4.5\times 10^{33}$erg\,s$^{-1}$.  

\subsection{Data reduction and processing}

We made use of the set of 37 X-ray pointings of the $Chandra$ ACIS-I camera that were
acquired in the context of the {\it Chandra} Cygnus OB2  Legacy Program (Drake et al., this issue).
An additional set of  three existing observations were included within the survey area, which had 
previously been used to study the Cyg~OB2 association \citep{Butt2006, Albacete2007, Wright2009}.
The total of 40 pointings were acquired in the VFAINT (5x5 pixel island) mode, which is an optimal choice  for 
identifying and removing events that originate from the end-points of particle event tracks and that
cannot be removed using a standard grade analysis based on 3x3 islands\footnote{See 
further details in http://cxc.harvard.edu/Acis/Cal\_prods/vfbkgrnd/}. This observational setup is crucial 
to disentangling true X-ray diffuse emission photons from instrumental and background effects.
The survey was performed over a central 0.5 deg$^2$ of the Cygnus OB2 association with an effective 
exposure of 120~ksec and an outer 0.35 deg$^2$ area with an exposure of 60~ksec. 
The description of the survey design and observations, the data reduction and
source detection, has been  presented extensively in \cite{Wright2014}.

All 40 ObsIDs were uniformly reprocessed using version 4.8 of the CIAO software 
\citep{Fruscione2014}, combined with CALDB 4.7.1 calibration database files. 
In order to get consistency in the calibration procedure, we re-ran the Level~1 
to Level~2 processing of event files using the CIAO {\sc chandra\_repro} meta task.
The {\sc check\_vf\_pha=yes} option was set to flag bad events and then filter them.
The new Level~2 event file was produced by filtering events to status==b0. 
The initial set of observation were processed with older {\it gain} files and were updated 
during the {\sc acis\_process\_event} to the lastest available file version.  Similar
treatment was applied to the background event files (see section 2.1 for details).
Since the available ``stowed" background files (from which the energetic particle event 
background can be estimated) were made using only quiescent background periods, 
background flares needed to be excluded from our observations in the exactly 
same way for consistency. We extracted light-curves
with the same time bin size as was performed for the background files (bin=259.28 sec), and we
made use of the {\sc deflare} CIAO\,4.8 Python 
routine\footnote{See http://cxc.harvard.edu/ciao/ahelp/deflare.html}. 
We time-filtered all of the observations.   The time reduction was found to be necessary, and 
in the worst case this amounted to less than 4\% of the total exposure of an observation.
In order to avoid intense instrumental emission lines (see section\,\ref{bkg}), we 
filtered the whole set of observations to exclude all events outside of the 0.5-7.0 keV energy range.

\begin{figure}[ht!]
\centering
\includegraphics[width=8.6cm,angle=0]{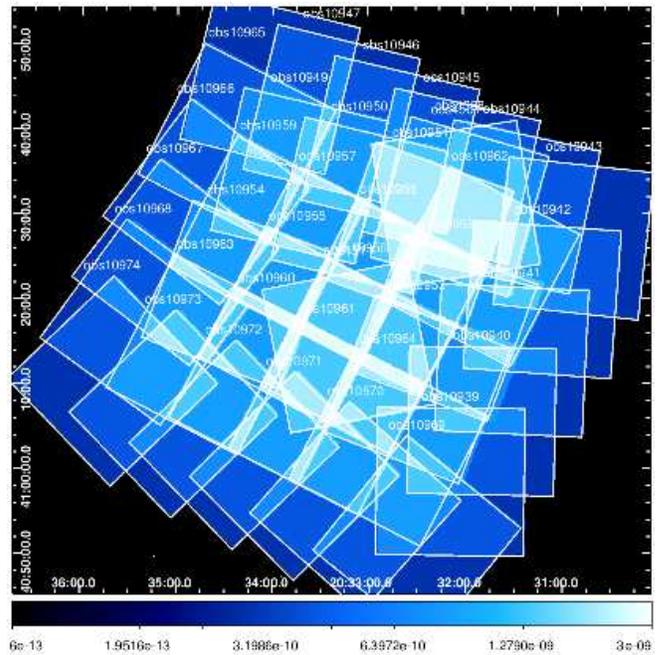}
\caption{\small The exposure map for the Cygnus OB2 Chandra Legacy 
Survey and complementary observations for the study of diffuse X-ray emission. 
The bottom color bar shows the flux to color  normalization factor in erg~cm$^{-2}$~s$^{-1}$.}
\label{expmap}
\end{figure}

The reprocessed event files and this set of observations 
are essentially the starting point of our analysis.
We computed monochromatic exposure maps for all 40 X-ray observations using the 
{\sc ae\_make\_emap} AE task for representative photon energies of 0.7, 1.7, and 3.5 keV. 
In Figure\,\ref{expmap}, we show the mosaic of exposures at 1.7~keV of all the observations used in 
the Cyg\,OB2 survey.  We use each of these sets of images to normalize the count images  
 in different energy bands of  [0.5--1.2], [1.2--2.5] and [2.5--7.0] keV, respectively, to produce X-ray fluxed 
images uncorrected for interstellar absorption.
However, as the diffuse X-ray emission is faint, the confirmation of its existence 
depends on careful analysis of the background (sub-section\,\ref{bkg}), 
as well as a thorough assessment of the contribution from X-ray point sources (sub-section\,\ref{mask}).

\subsection{Background analysis}
\label{bkg}

\begin{figure}[!h]
\centering
\includegraphics[width=8.4cm,angle=0]{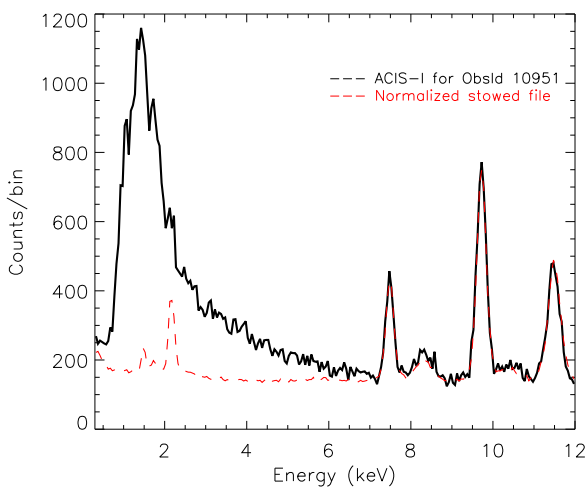}
\caption{\small Pulse height distribution of events for ObsId 10951 (black continuous line) 
together with the stowed background normalized to match based on fluxes at energies of  
7.47, 9.8 and 11~keV, where the sky has no emission (red dashed line). 
The weaker $\sim$ 2.1 keV emission line is also instrumental, but was not used to match
stowed-observation events and so it could be present in some diffuse X-ray spectra.}
\label{depht}
\end{figure}

Analysis of faint extended X-ray diffuse emission is an inherently difficult  task, as the emission 
is spread over the detector and is affected by non-local 
background such as solar and radiation belt energetic particle events, and X-ray events from 
interloping sources such as active galactic nuclei (AGNs).  
As noted above, the instrumental and energetic particle background was assessed using a set of ACIS stowed 
observation files\footnote{http://cxc.cfa.harvard.edu/contrib/maxim/acisbg/data/}. 
Since the ACIS instrumental background is known to be time-varying, the proper 
scaling of these images to a given observation cannot be estimated from relative exposure times 
alone (see, e.g., details in Section 4.1.3 of \citealt{Hickox2006}). 
Thus, background images were also scaled to match the spectra of each of the 40 observations in the 
7--12~keV range, where the stellar emission has no significant signal (see Figure\,\ref{depht}).

Finally, we constructed new stowed exposure maps for the next step of the analysis, which 
uses mask stowed event files and new stowed exposure maps for use at each source  position.

\subsection{Masking X-ray point sources}
\label{mask}

\begin{figure*}[ht!]
\centering
\includegraphics[width=8.4cm,angle=0]{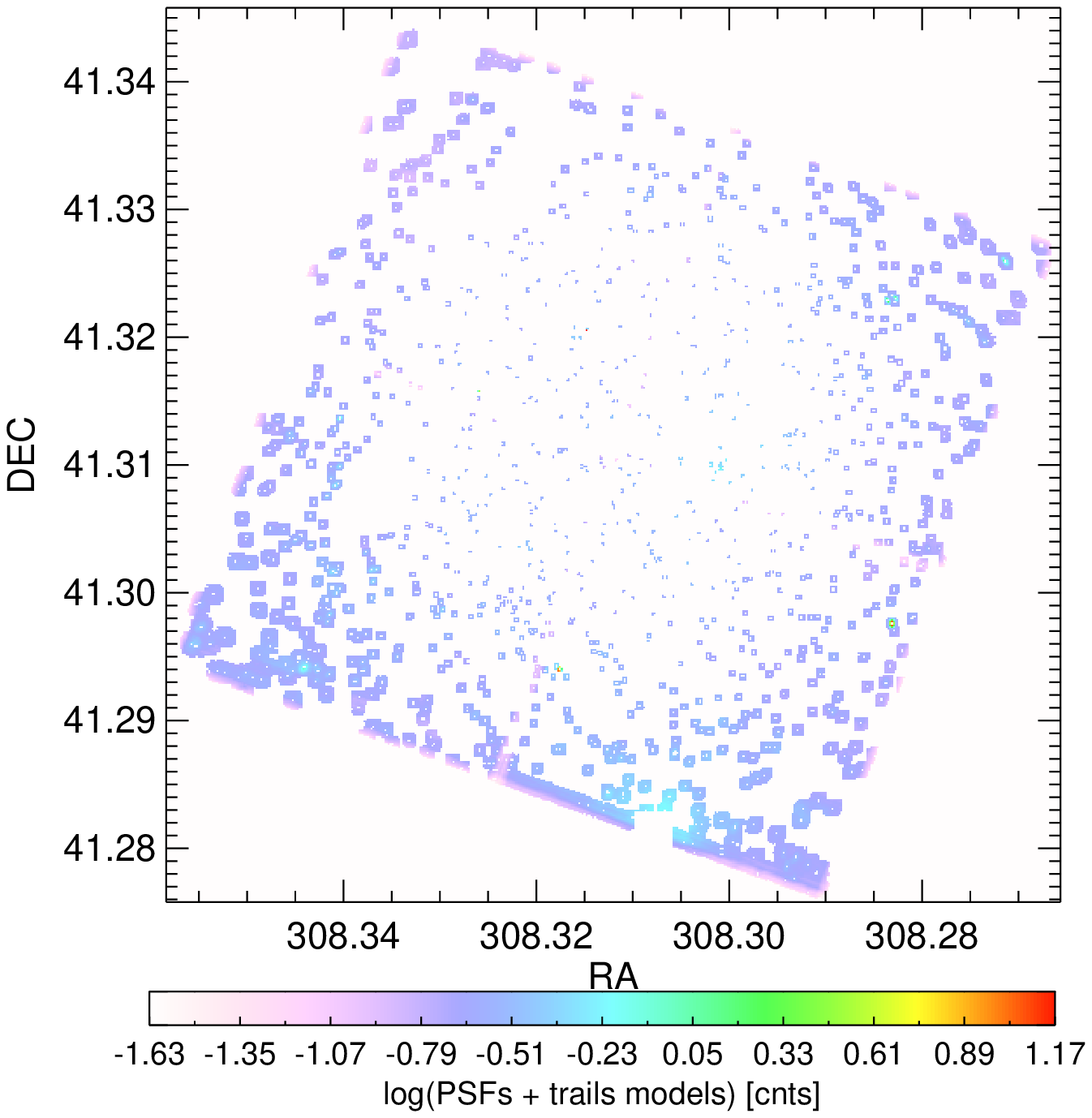}
\includegraphics[width=8.4cm,angle=0]{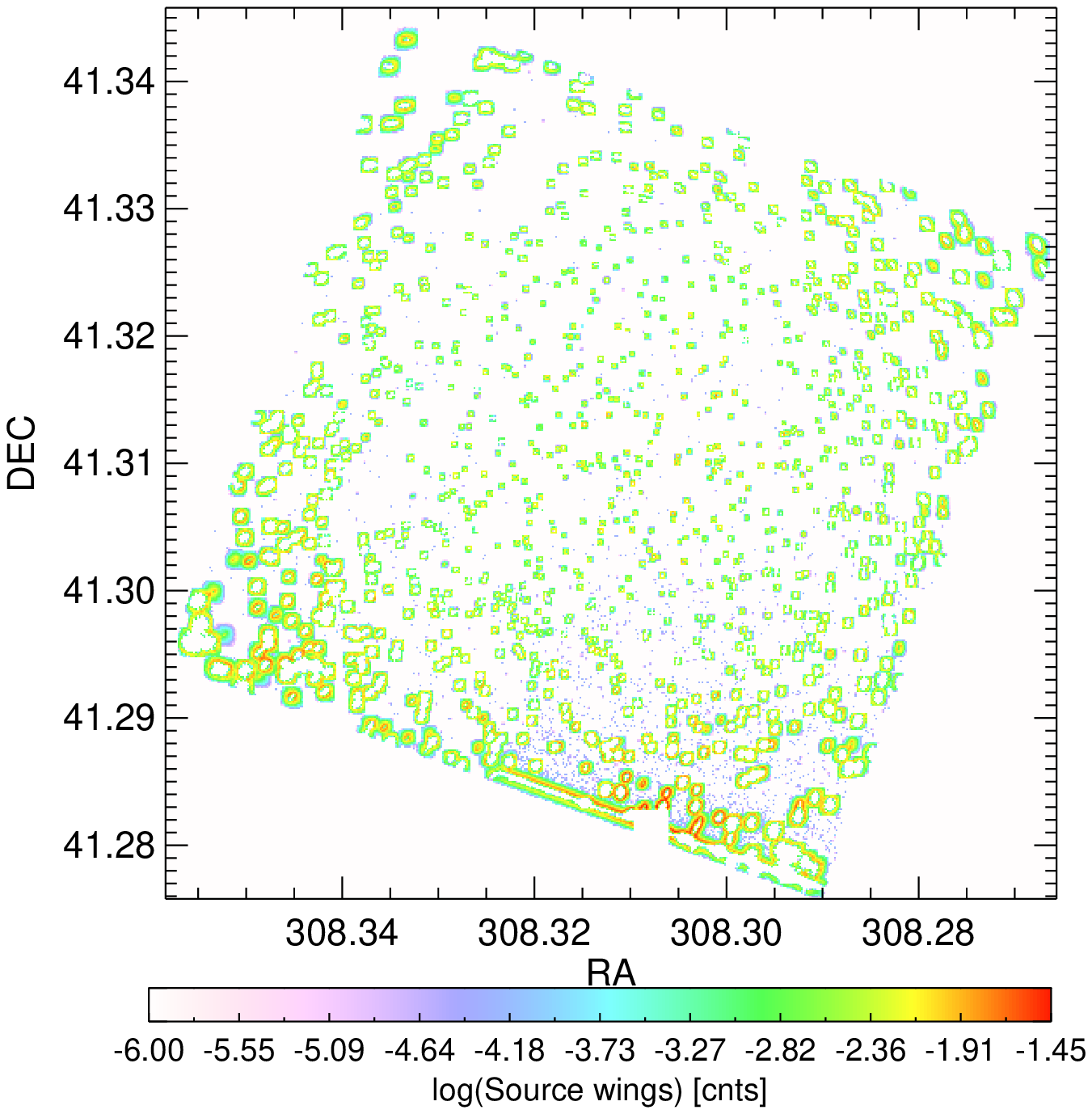}
\caption{\small ObsId 10951, illustrating the treatment of point sources and their exclusion 
from diffuse emission maps.  Same analysis were applied to the whole set of observations. 
Left: Show the source PSF and readout streak intensity models that
were computed on the list of detected sources in the observation.
Right: Show the residual PSF wings image which was computed as the difference between 
the observed source count image and the source+streak intensity models (left-image).
Scale bars of both panels show the intensity range in which the images spans.}
\label{wings}
\end{figure*}

Here we give details of the procedure implemented to reduce the impact of X-ray point sources 
on the event and calibration (stowed) data files.  We used 
the specific task {\sc ae\_mask\_stowed\_data} from the AE code that masked event files, 
exposure maps, and both stowed events and stowed exposure 
files. We thus subtracted events from a total of 7924 X-ray point sources listed in 
our catalogue \citep{Wright2014} covering all 40 of the ACIS-I observations.

\begin{figure*}[ht!]
\centering
\includegraphics[width=5.8cm,angle=0]{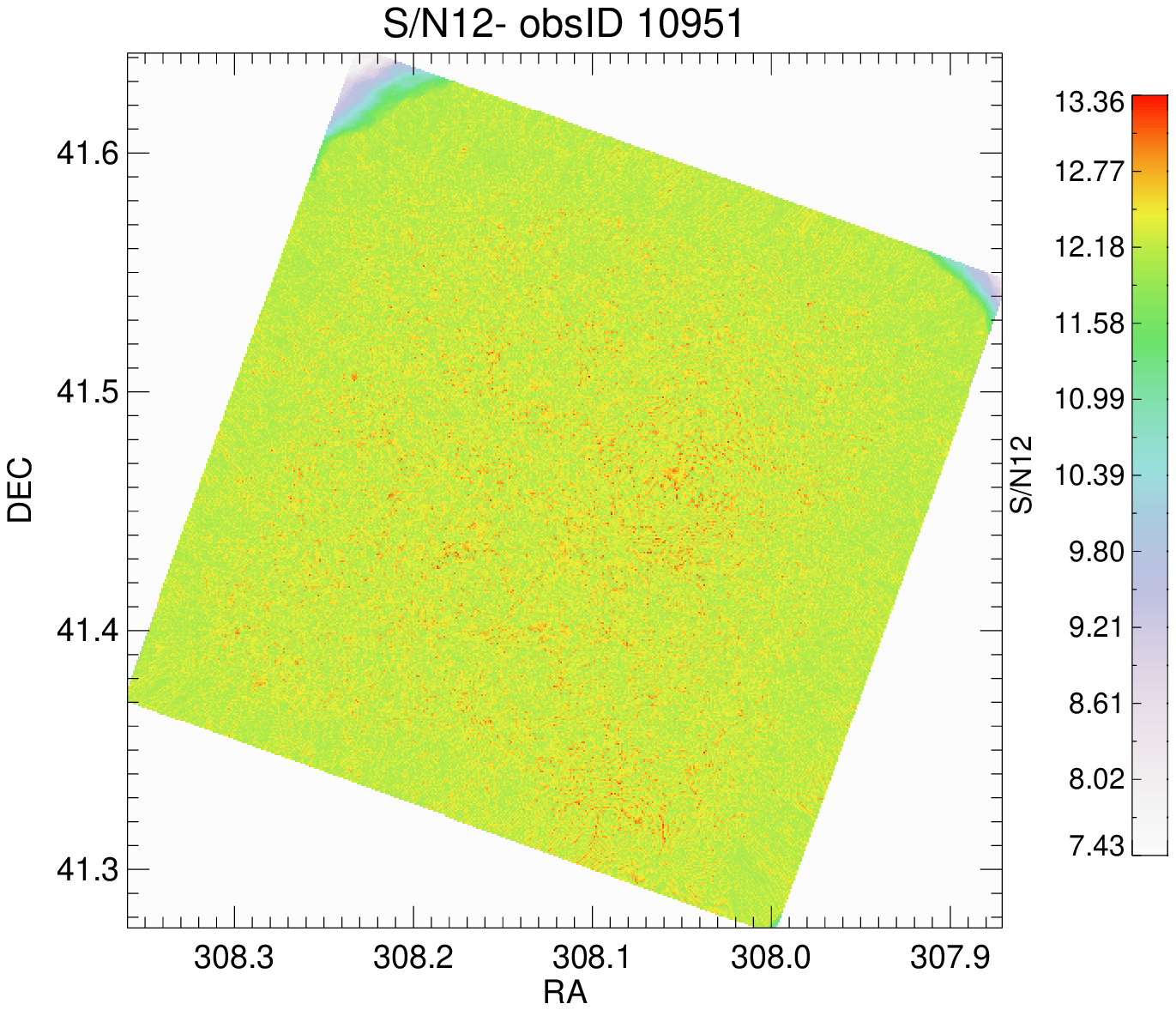}
\includegraphics[width=5.8cm,angle=0]{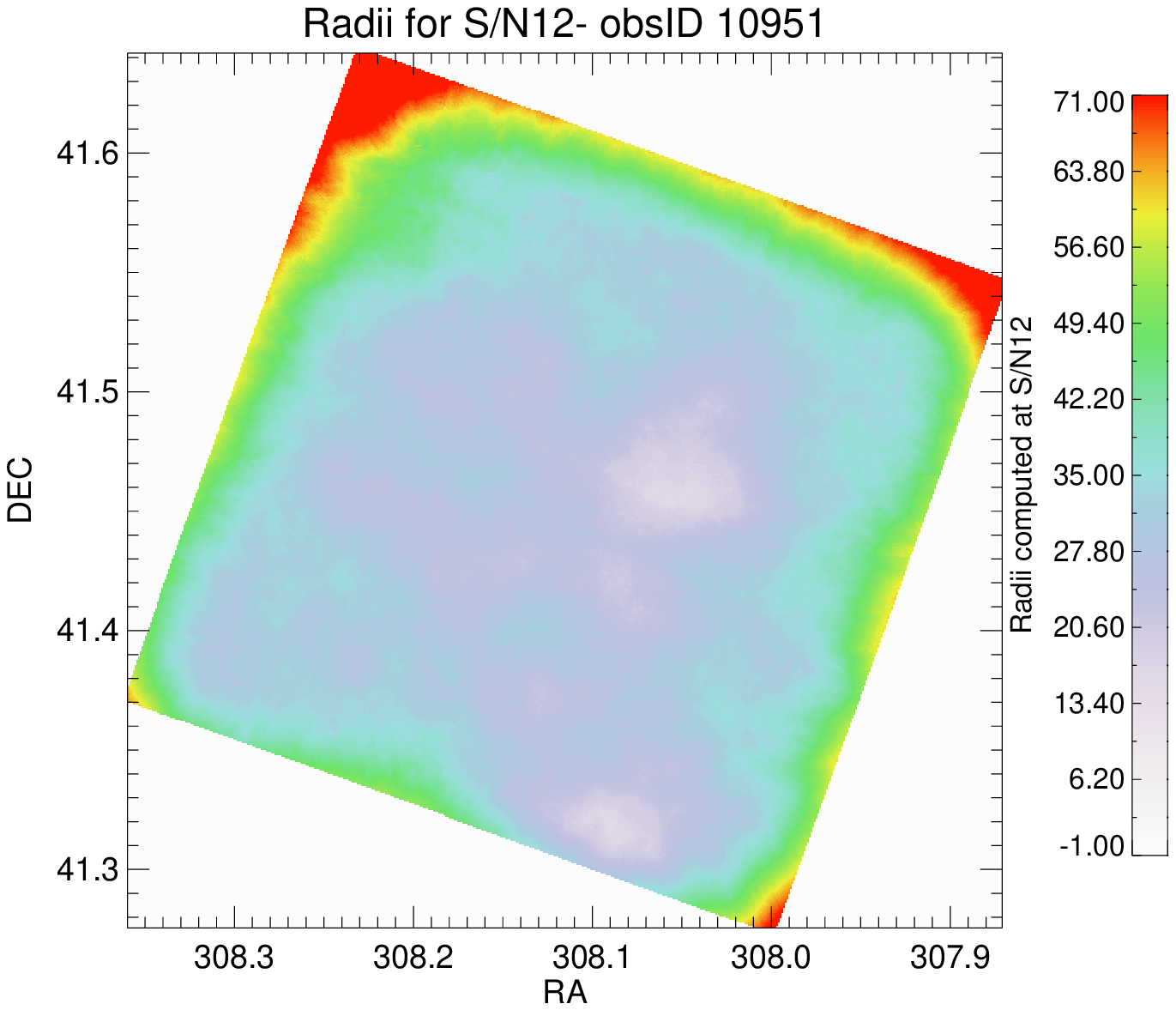}
\includegraphics[width=5.8cm,angle=0]{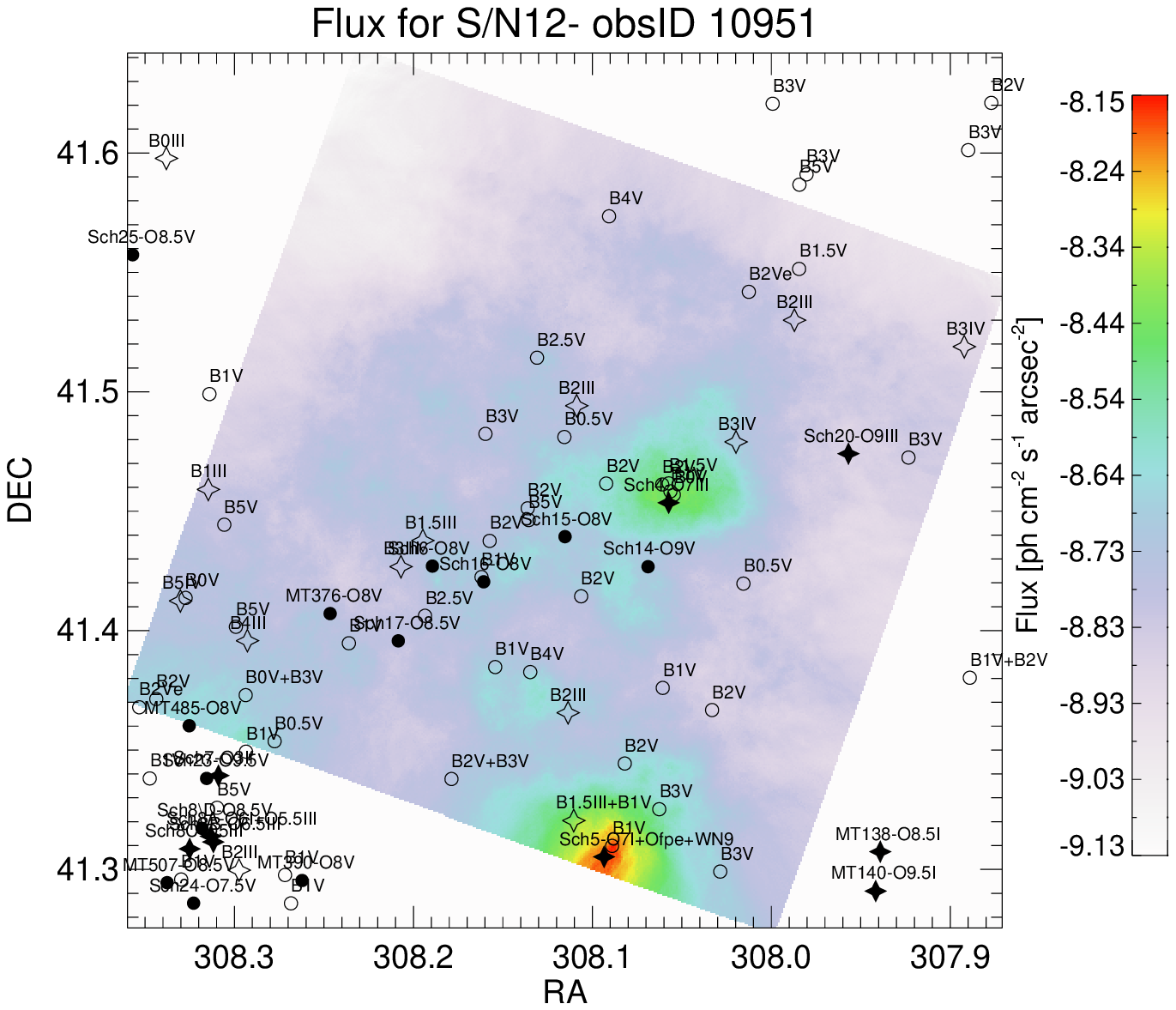}\\
\includegraphics[width=5.8cm,angle=0]{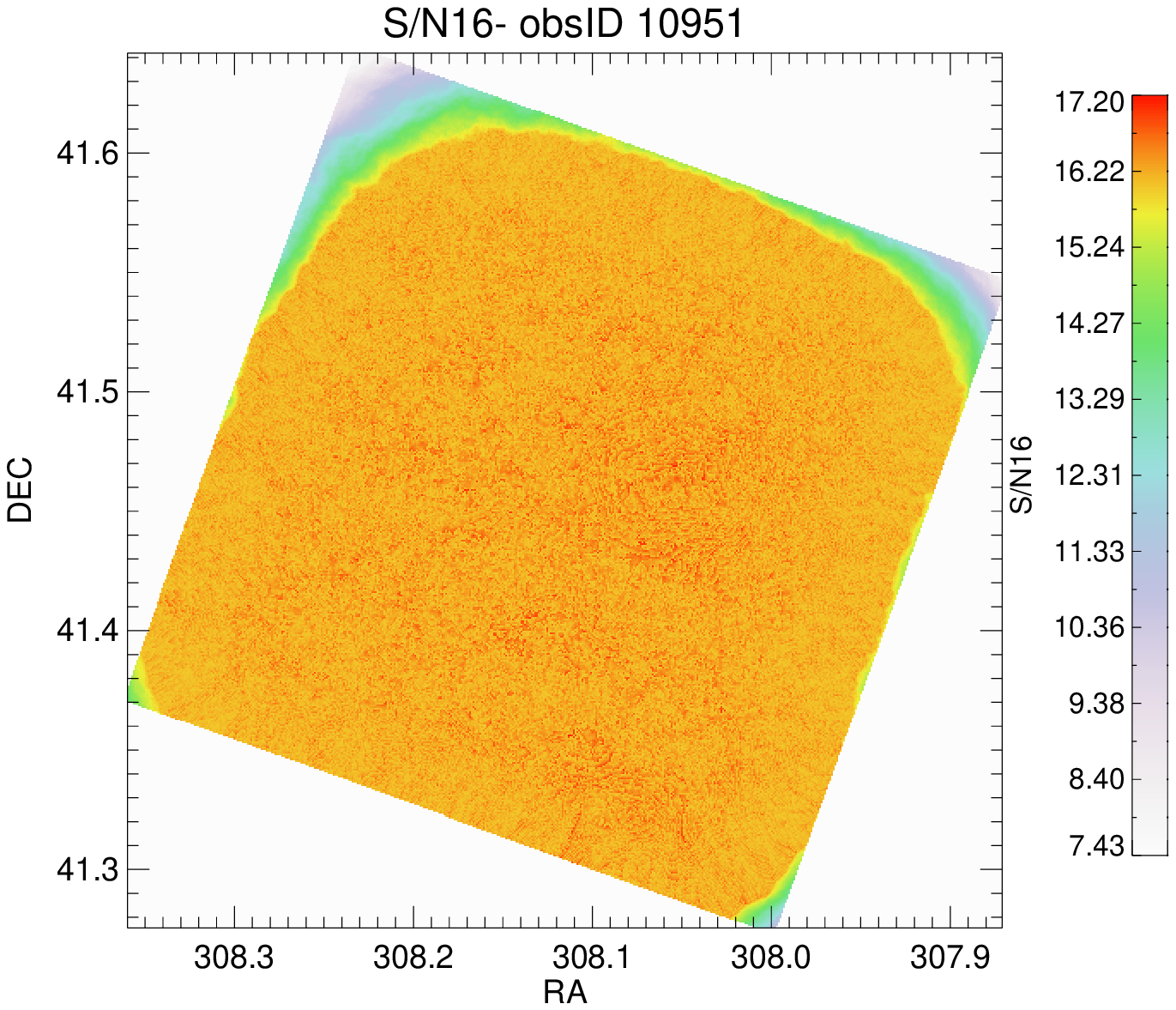}
\includegraphics[width=5.8cm,angle=0]{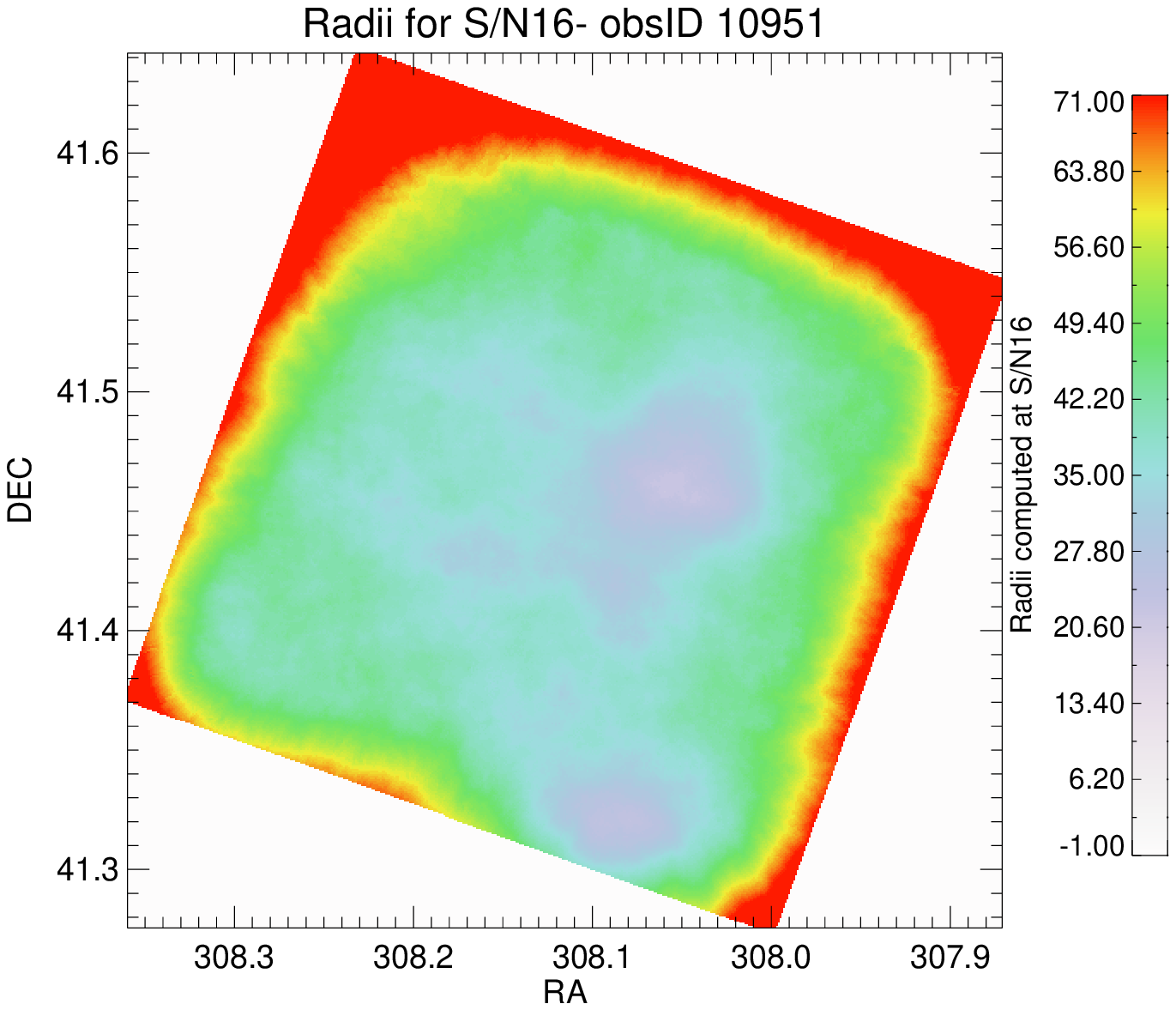}
\includegraphics[width=5.8cm,angle=0]{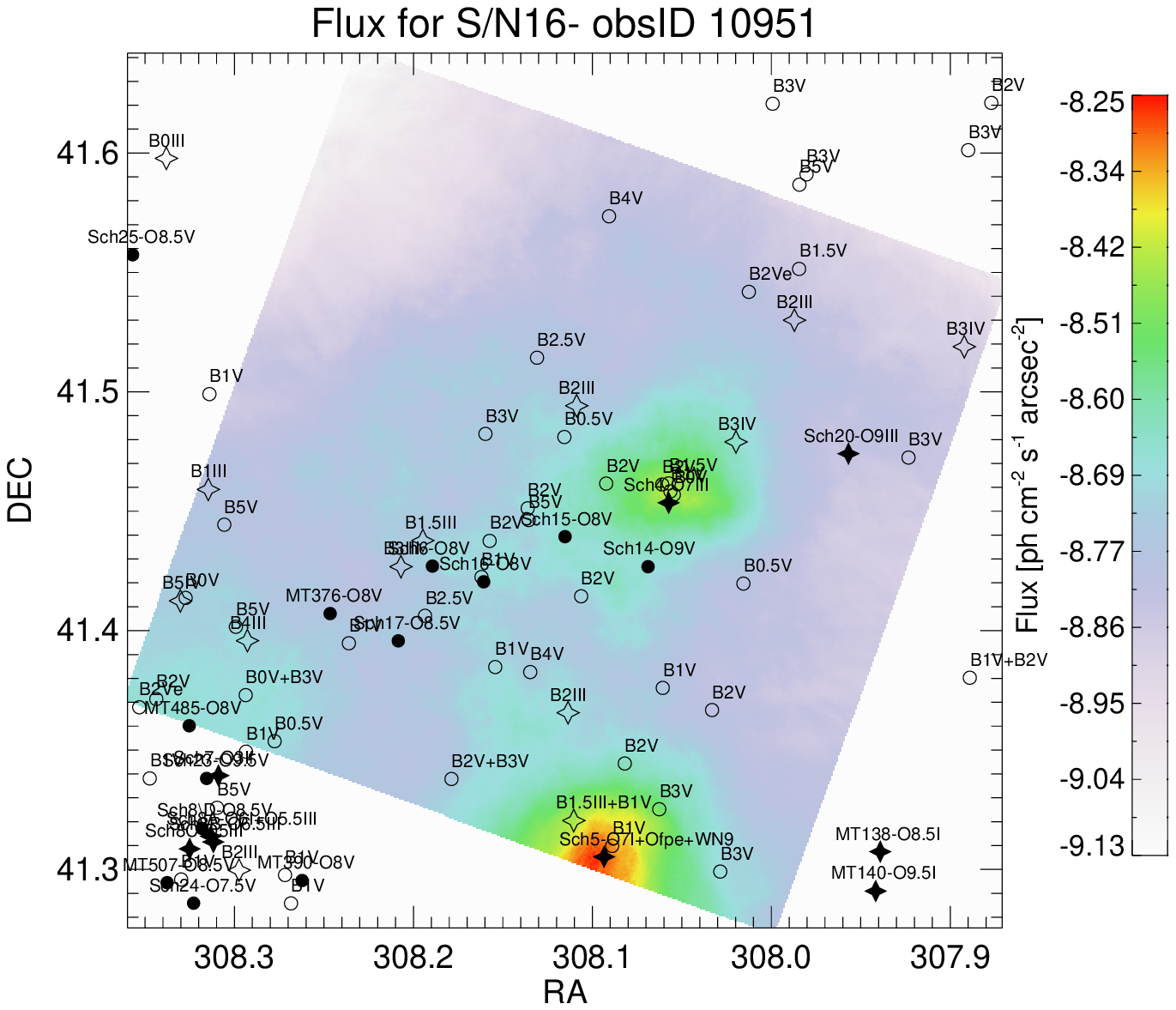}
\caption{\small Exemplification of the procedure for the ObsId 10951. 
Left panels: the S/N significance map computed at the imposed condition of S/N=12 and 16 respectively. 
 Central panels: The tophat radius computed to reach the imposed S/N condition.
Up to the borders of the observation, the tophat radii increase as much as necessary to achieve the S/N, although once off the
detector the signal falls to zero and can reduce the final S/N below the desired threshold.
Right panels: X-ray diffuse emission flux [ph/cm$^{2}$/s/arcsec$^{2}$] in the [0.5--7.0] keV energy range 
computed at S/N$\sim$12 and 16, respectively. Note that the peak of emission changes slightly for different 
S/N owing to the different binning, but by just  0.1 dex or roughly  25\%. 
At the limit of detection of the diffuse emission of log(F$_{\rm x}$)$\approx$-8.64 or -8.69, for S/N=12 or 16 respectively, 
the difference is only 0.05 dex, or about 11\%.
Black filled star symbols indicate the evolved massive stars with
conspicuous X-ray emission; open star symbols refer to evolved massive stars without X-ray emission; 
black filled and empty circles indicate main sequence massive stars, with and without 
detected X-ray emission, respectively. The list of massive stars in Cygnus OB2 was taken from \cite{Wright2015a}.}
\label{flux}
\end{figure*}

\begin{figure*}[ht!]
\centering
\includegraphics[width=8.3cm,angle=0]{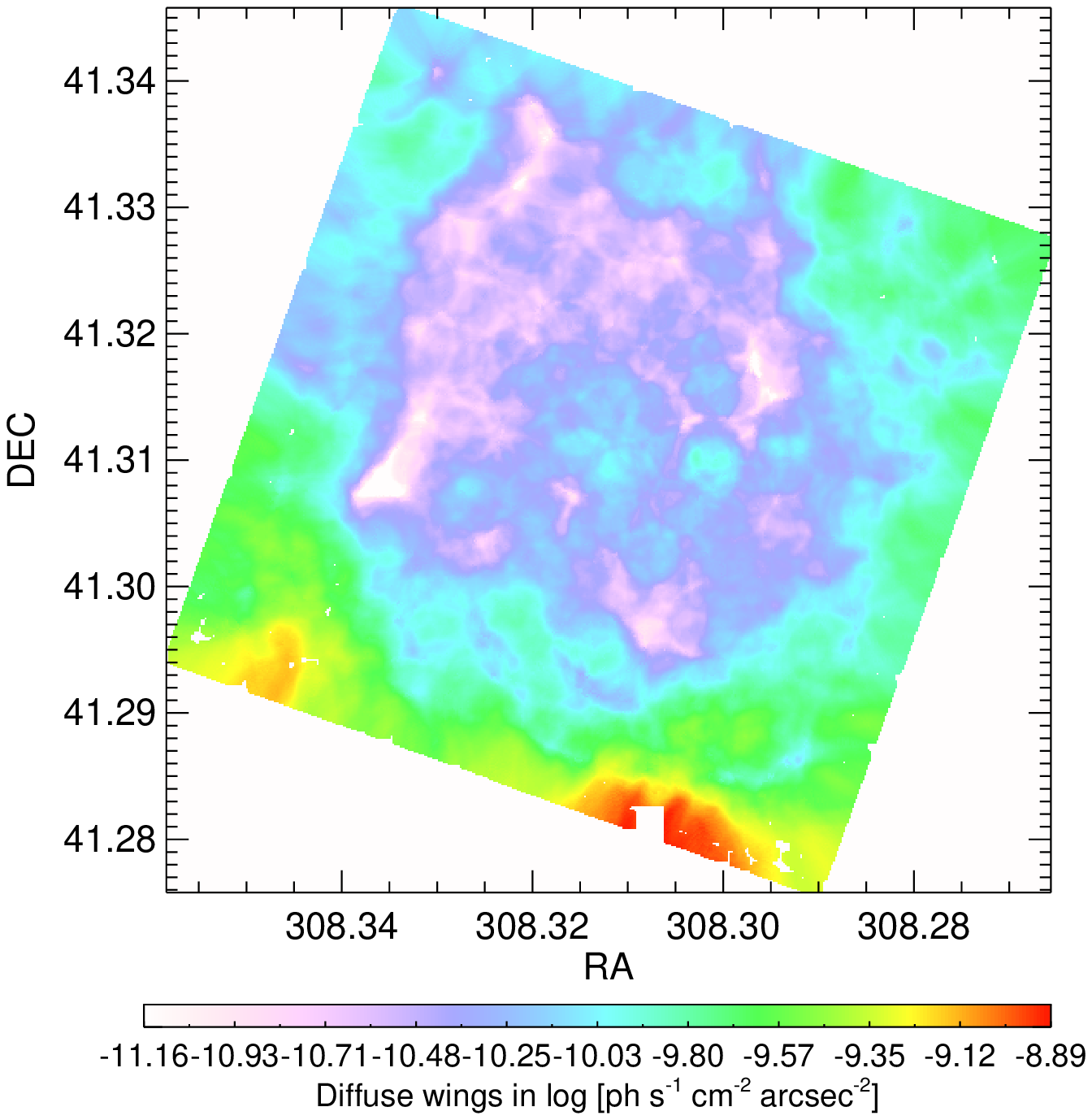}
\includegraphics[width=8.3cm,angle=0]{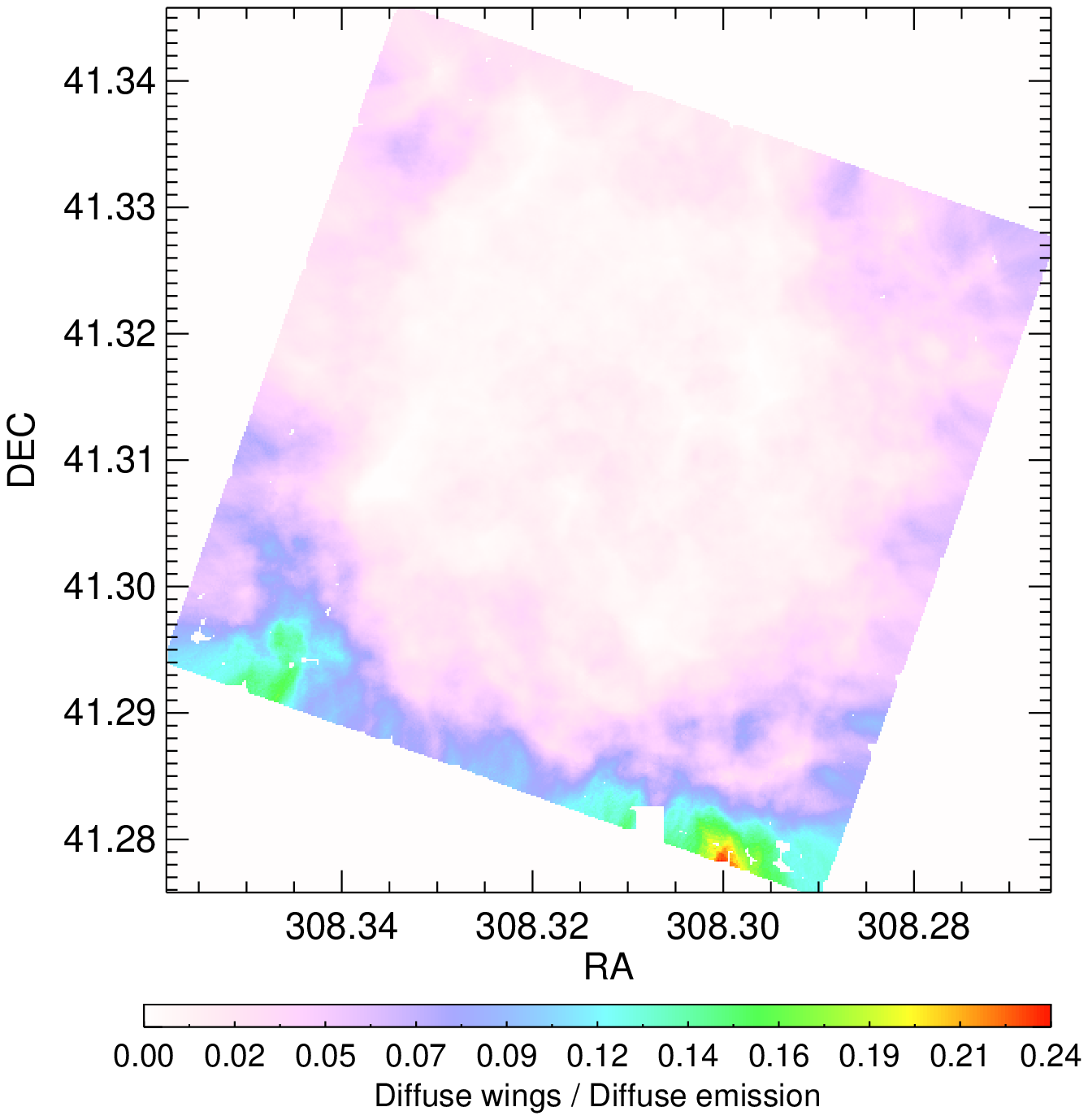}
\caption{\small Exemplification of the procedure for the ObsId 10951. Left: Source wing diffuse contribution in the [0.5-7.0]~keV 
energy range computed using fixed smoothing radii for S/N$\ge 16$ in the full band. 
Right: the ratio of wings to total diffuse emission has a maximum contribution of 0.24,
which lies at the border of the map. The median contribution from wings all over the image
is about 6\%.}
\label{wings-test}
\end{figure*}

The first step in the point source subtraction is to build intensity models for sources, as well as for readout streaks, 
and draw masks around them to compute an image that models the signal from all these features.
We use the task {\sc ae\_better\_masking} from AE to adaptively compute source emission 
at the 99\% enclosed energy aperture threshold. Such a mask fraction is more than adequate 
for faint sources with only a handful of counts, since the probability 
of losing genuine source events in the PSF wings is low.
In cases where the X-ray source was intense, the source wings were inspected close to the masked 
event data to see if there were remaining wings from scattering photons that are bright enough 
to locally contaminate our diffuse emission analysis. 
In such cases, we arbitrarily increased the mask size by multiplying up to a factor 1.5 times 
the 99\% PSF limit \citep{Townsley2011a}. Under this condition, the AE recipes assure that source 
event photons in the wings of the PSF, or photons scattered out of the PSF, 
should fall below the local background level of the observation \citep{Broos2012}.
In Figure\,\ref{wings}-left, we show the resulting source\,+\,readout streak emission models.
Otherwise, the Figure\,\ref{wings}-right show the residual map of the cropped sources,
that consist in the difference between the observed source count image and 
the source+streak intensity models (left-image). Hereafter we refer to these residual images as "wings" files. 
Note that the wings image peaks at the log(Source wings)$\approx$-1.45. It means, that 
the maximum ''wings'' contribution to the observed emission image is only about 3.5\%, 
or less. 
Both maps shown correspond to obsID\,10951, that hereafter will be used to illustrate 
the analysis that was applied to the rest of the observations.\\

\section{Adaptive smoothing strategy}
\label{tophat}

One of the most critical issues in the study of X-ray diffuse emission is the choice of smoothing  strategy. 
For this purpose, we made use of the {\sc tara\_smooth} 
tasks\footnote{These tasks are not part of the public AE software, 
and were kindly shared by Dr. Patrick Broos of Penn State University's astrophysics group.}. 
We used the {\rm top hat} adaptive kernel smoothing.
All maps were computed in a 512$\times$512 array. Larger 1024$\times$1024 maps did not produce 
better results and were also computationally very demanding.
Two main parameters that play a major role in smoothing are the i) significance, which is 
a scalar or vector number specifying the desired Signal-to-Noise (S/N) ratio of the smoothed flux image, and 
ii) the smoothing  radii that could be imposed to achieve the same spatial smoothing scales in different
energy bands. The radii of smoothing are limited to a maximum of 71 pixels, and the imposed S/N condition is only applied
if smoothing radii remain below this limit.
The smoothing procedure was initially run for the full [0.5--7.0] keV band.
After several runs at S/N ratios of 9, 12, 16, and 25,  
we found that S/N$\sim$16 is the best compromise between  the imposed S/N condition,
 smoothing radii, and the ability to unveil true X-ray diffuse structures at spatial scales 
 greater than the smoothing radii all over the field of view (see Figure\,\ref{flux}).
The units of the resulting flux maps  are ph/cm$^{2}$/s/arcsec$^{2}$.

For sub-band images in the energy ranges in keV
Soft [0.5--1.2], Medium [1.2--2.5], Hard [2.5--7.0] and total [0.5--7.0],
we adopted a different smoothing strategy.
We decided to smooth sub-images using the same set of radius kernels computed to 
achieve a given S/N ratio in the band with poorest photon statistics. 
We adopted S/N$\sim$16 to avoid smoothing radii becoming too large, with this condition relaxed at the borders of the detector.
For the rest of the bands, smoothing was performed by fixing the smoothing radii map to that of the Soft band via 
the optional parameter {\sc fixed\_radius\_map\_fn}. As the Medium and Hard X-ray bands 
have  better photon statistics, the significance of such
maps reach higher S/N than in the soft band.
In  this way, we guarantee that the adopted S/N ratio is always achieved for the rest of the energy band 
maps, as well as ensuring that the spatial scales of the maps are adequate for construction of hardness ratio maps.

\subsection{The X-ray source wing diffuse contribution}

Based on the smoothing  considerations presented in Section\,\ref{tophat}, we are able to
estimate the contribution to diffuse emission produced by photon events in the extreme 
wings of sources that were not  adequately excluded by the computed masked regions.
To do this we applied the same smoothing constraints used for the diffuse maps to the
source wing diffuse maps of Figure\,\ref{wings} (right). 
The {\sc tara\_smooth} routine was fixed to radii computed for the flux diffuse emission at S/N$\ge16$. 

\begin{figure}[h!]
\centering
\includegraphics[width=8.7cm,angle=0]{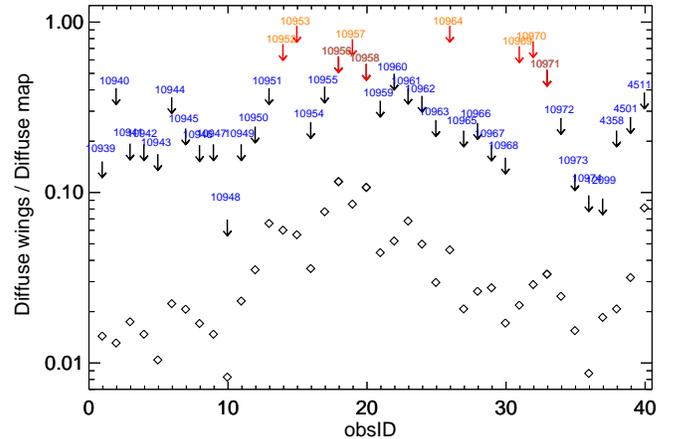}
\caption{\small Down arrows indicate maximum contributions of source wings to diffuse emission (see text for details).
Red arrows indicate observations that are likely to be affected by gas and dust scattering processes
related to the intense emission from Cygnus X-3.   
Black diamonds correspond to the median of the wings to diffuse 
emission ratio for the entire FOV of each observation.}
\label{contam_ratio}  
\end{figure}

 In Figure\,\ref{wings-test} (left) we show the resulting diffuse wing emission 
contribution for ObsID 10951. The emission peaks at $\log({\rm f}_x) =-8.89$~ph/cm$^{2}$/s/arcsec$^{2}$,
at large off-axis angles as a consequence of ill-constrained source\,+\,readout streak emission models 
related to the intense diffuse X-ray stellar source Schulte\,\#5 (see section 3.1).
In order to quantify such possible contributions over the entire set of the observations, 
we computed the diffuse wings\,/\,total emission (f$_{\rm wings}$ / f$_{\rm diffuse}$) 
maps to identify possible spatially-resolved wing contamination zones.
In the right panel of Figure\,\ref{wings-test}, we can assert that, except for the very edges of the observation field of view, 
diffuse wing source contamination remains, in media, under 6\%\ over the FOV.

 Next, we computed the median ratio of the f$_{\rm wings}$ / f$_{\rm diffuse}$ contamination
for each of our observations. In media, these values illustrated in Figure\,\ref{contam_ratio}, 
range between 1 and 10\,\%.
However, there are some regions in which the ratio f$_{\rm wings}$ / f$_{\rm diffuse}$ 
peaks at over 50\% (ObsIds 10952, 10953, 10956, 10957, 10958, 10964, 10969, 10970 and 10971). 
This occurs only for some intense sources located at large off-axis angles, or at the chip borders 
of the observations because of read out streak events or because the source PSF itself cannot accurately 
modelled. 
These high ratios do not inconvenience our diffuse analysis at all, as they correspond to very
small fractions (typically $\le$ 4 to 6 \%) of the map areas. 
We avoided these parts of observations in the diffuse X-ray analysis, and 
instead used observations in which the same sky region appears close to on axis,  
where the masked model describes adequately the local PSF of the sources.\\

\section{X-ray hardness ratio maps}

The construction of X-ray hardness ratio (HR) maps is a useful method of getting a first
order approximation of the spatial energy distribution of X-ray photons without losing
spatial resolution. We produced [count-rate] maps in (S)oft [0.5--2.5 keV] (S) and (H)ard 
[2.5--7.0 keV] X-ray bands, and computed HR as the ratio of the difference to the sum, i.e.
(S-H) / (S+H). In this way we are able to discern soft and/or hard features in the diffuse X-rays 
regardless of their relative intensity.
All HR maps were computed from maps that were smoothed using the {\rm tophat} method, but 
considering the fixed smoothing radius computed for a signal-to-noise $\ge$16 in the soft band
(see $S\ref{tophat}$ for a full explanation). We thus avoid false diffuse structures that would appear 
due to differences in the radii of the maps.
HR maps were constructed for the entire set of observations  in the survey (see Appendix).
\label{hrsection}

\begin{figure}[h!]
\centering
\includegraphics[width=8.9cm,angle=0]{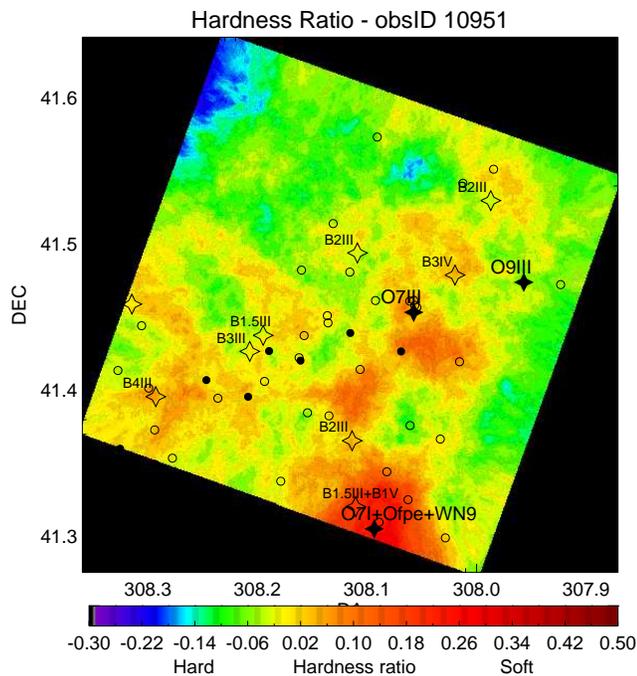}
\caption{\small Exemplification of the HR analysis for the ObsId 10951.
The scale color bars show the range in HR that the image spans. 
Black filled and opened star symbols indicate the evolved massive stars with and without 
intrinsic X-ray emission, respectively (from \cite{Wright2015a}).  Filled and open circles indicate 
MS massive stars, with and without intrinsic X-ray emission, respectively.
Note that the diffuse gas follows the spatial distribution of the massive stellar population, 
regardless of whether they are strong X-ray emitters. The entire set of HR maps is shown
in the Appendix.}
\label{hrsingle}
\end{figure}

Figure\,\ref{hrsingle} shows that softer diffuse emission follows 
the spatial distribution of massive stars from \cite{Wright2015a}, even 
in those cases where massive stars do not emit significant X-rays themselves. 
As we discuss further below,  the sensitivity of the HR maps highlights 
small changes in the energetics of the diffuse X-ray emission, providing clues 
about the impact of massive star stellar winds on the
spectral energy distribution of the diffuse X-ray gas, even in those regions 
bereft of bright X-ray sources or with high ISM density.

The natural explanation for what we are seeing is that evolved massive stars in the region fill the 
volume of space between the massive stars with the summed contribution of hot 
shocked stellar winds. 
Eventually, these encounters drive a slow shock into the ISM, that contributes to the 
excitation of the observed H$_\alpha$ emission and, via the presence of dense neutral gas 
and/or dust structures absorbs and re-emits radiation in X-rays
(see section\,\ref{discussion} for discussion).

However, not all HR values in the maps are descriptive of the energy of
diffuse X-ray emission from shocked gas, and 
 in those places where the diffuse emission is undetected, or absent, 
the extragalactic X-ray background also appears to play a role in some cases (see sub-section\,\ref{hrbgk}).

\subsection{Hardness ratio of the background}
\label{hrbgk}

\begin{figure}[h!]
\centering
\includegraphics[width=8.5cm,angle=0]{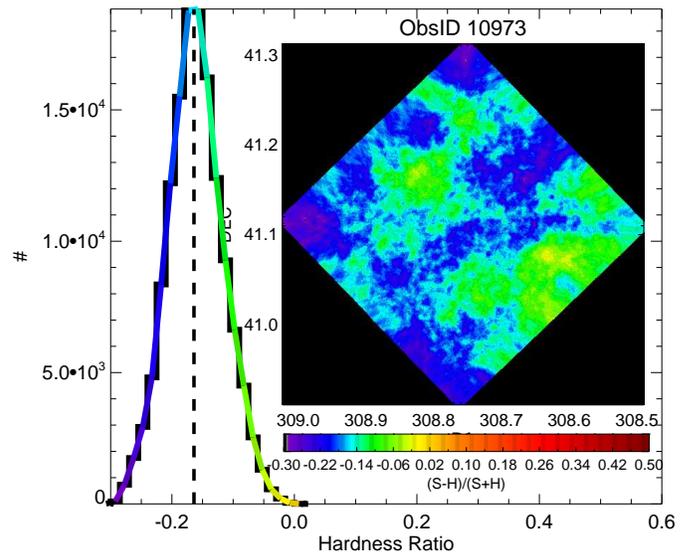}
\caption{\small Hardness ratio histogram for ObsID 10973. No obvious Cygnus~OB2 X-ray diffuse emission was
seen in this observation (see middle panel-right of figure 27 in Appendix), 
so that the observed HR corresponds to that of the background emission.
Colors in the histogram corresponds to those of the color bar of the image.
The peak of the HR distribution is about -0.16 for a 1$\sigma$ of 0.06.}
\label{hrhisto}
\end{figure}

\begin{figure*}[!ht]
\centering
\includegraphics[width=5.7cm,angle=0]{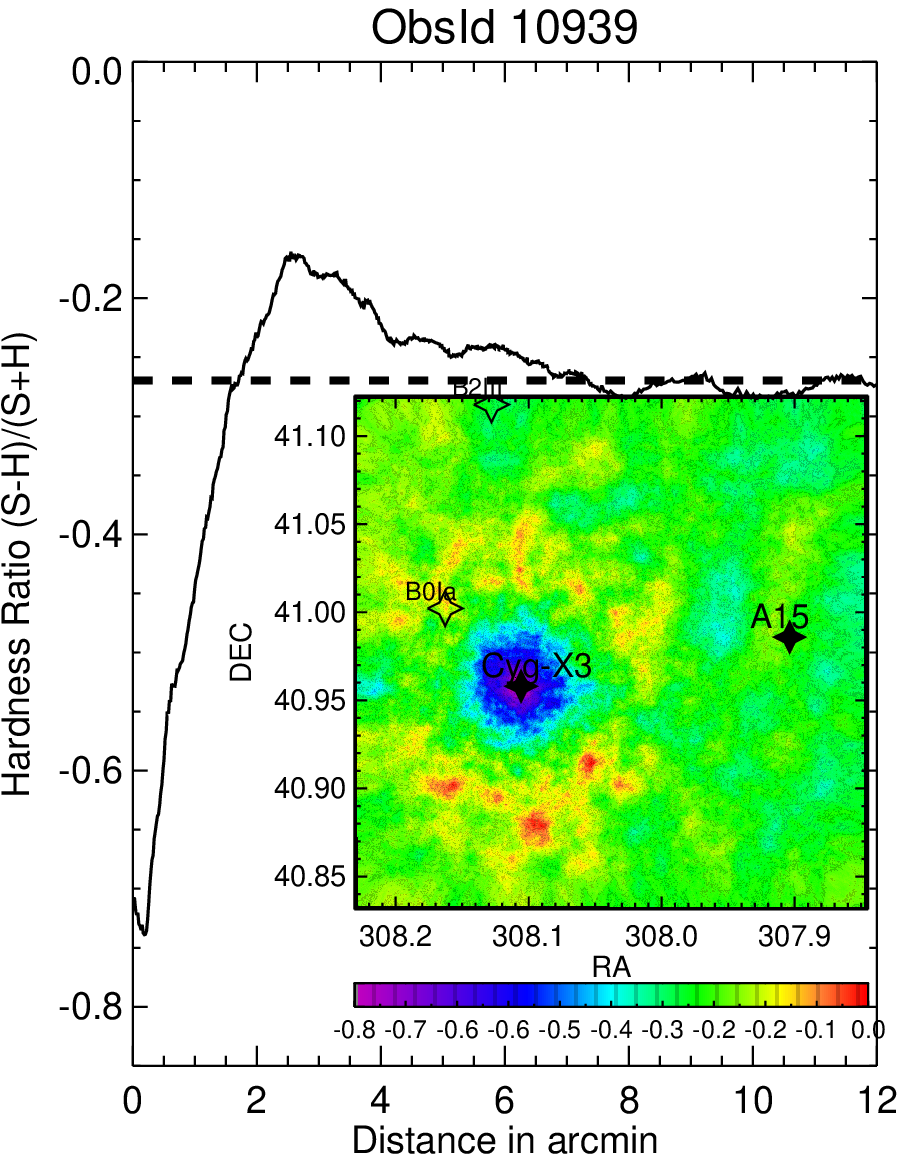}
\includegraphics[width=5.7cm,angle=0]{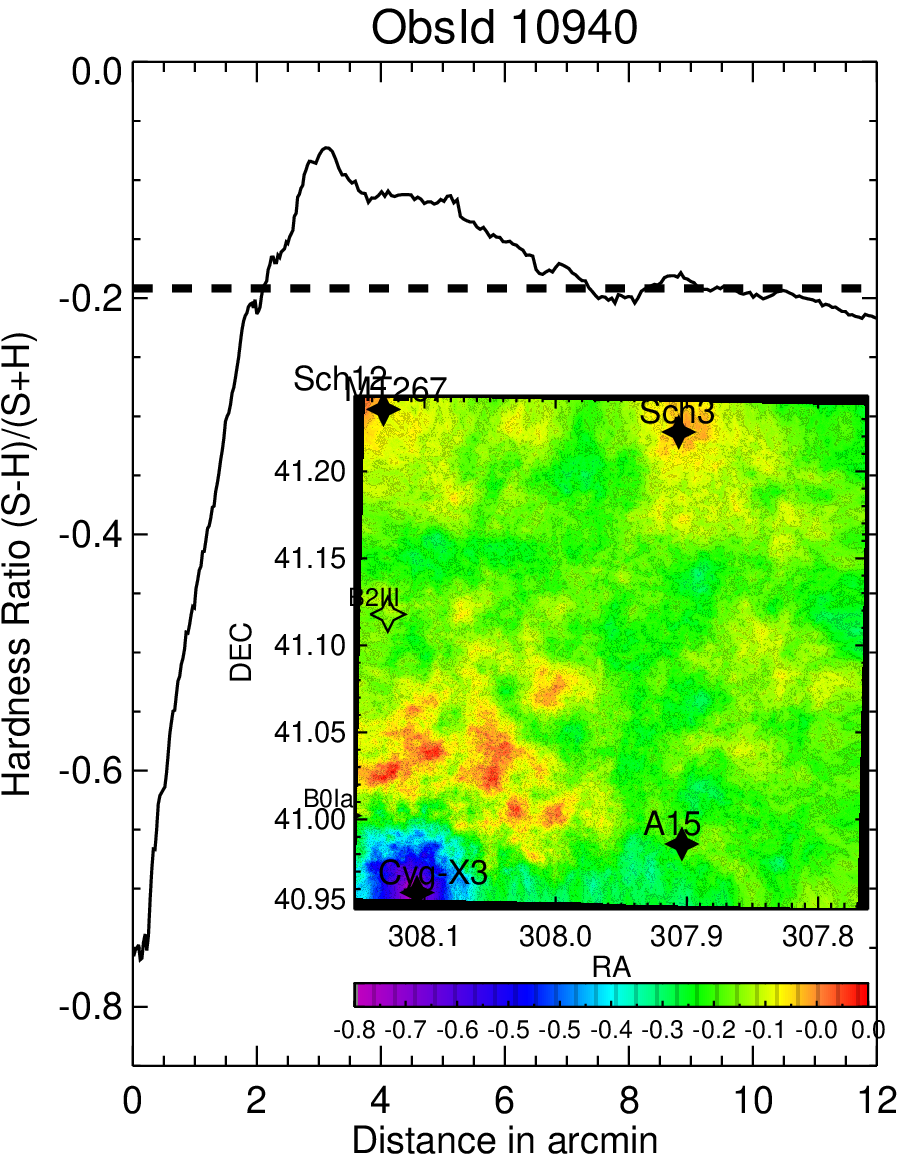}
\includegraphics[width=5.7cm,angle=0]{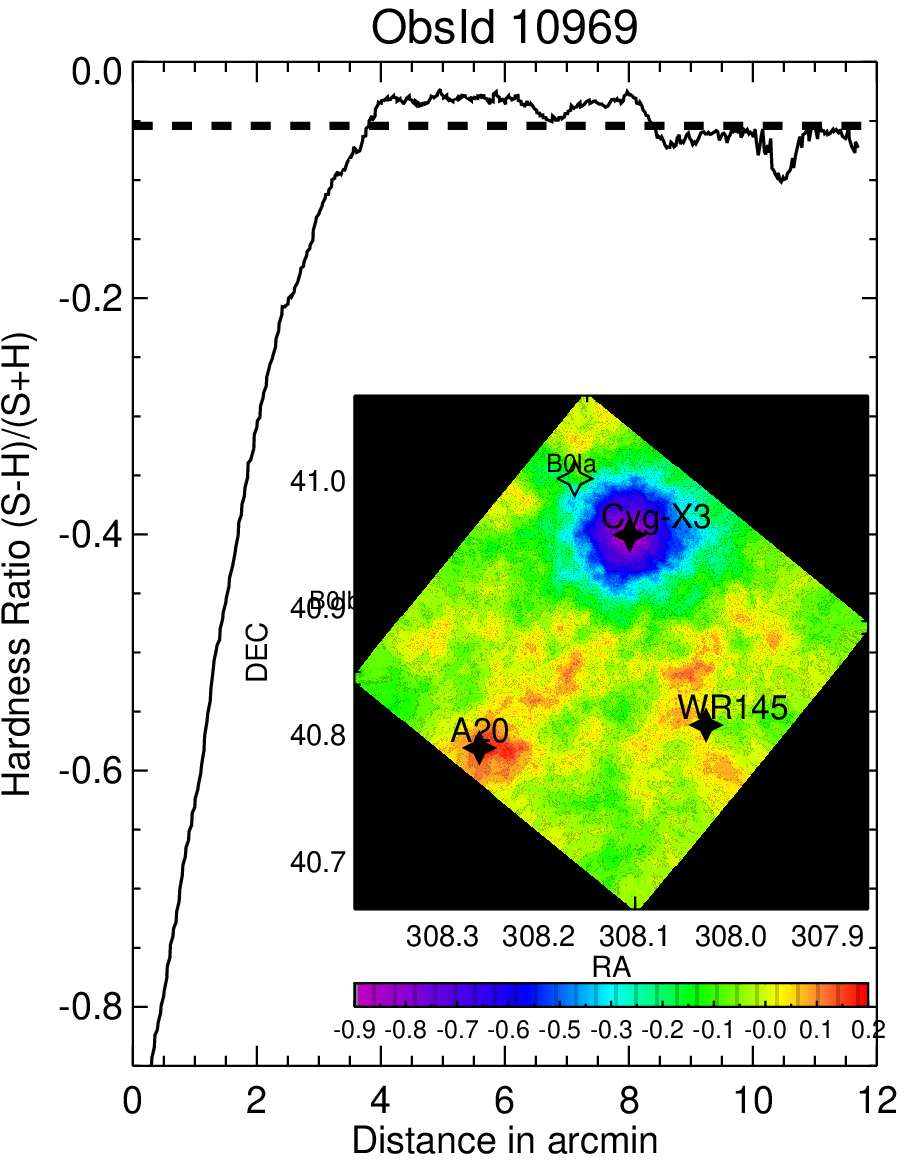}
\caption{\small  Cyg X-3 HR radial profiles of HR maps. 
Diffuse scattering extends to typical HR$\, \approx$ -0.14$\pm$0.07, where 
in worst of case, it drops to the level of the ambient 
Cyg~OB2 HR at a radial angular scale of $\approx$ 8 arcmin. Beyond this range we consider 
the influence of Cygnus X-3 scattering on the diffuse X-ray emission to be negligible.}
\label{radprof}
\end{figure*}

A serious difficulty in characterizing the diffuse emission energy distribution using HR maps 
concerns the problem of where the background becomes dominant, 
and how to estimate a typical HR value for it. 
We make an initial estimation by searching for X-ray observations (or places in Cygnus OB2) 
in which X-ray diffuse emission appears to be absent (e.g.\ ObsID 10973 in Figure\,26 of the Appendix). 
In this observation, no massive stars are in the  FOV, and it is also unaffected by 
background scattering of X-rays from Cygnus X-3.

We computed the HR energy distribution of the smoothed background 
for the entire field of view of obsID 10973, finding it peaks at -0.16$\pm$0.06. 
We have simulated a set of power-law emission spectra that adaptively constrains the observed 
HR values on this map. Typical $\Gamma$ indexes range from 1.0 to 1.3, which agree with the expected spectral 
index for Active Galactic Nuclei (AGN). In fact, this estimate of the hard X-ray background 
is known as the "X-ray Background Hardness Problem" and originates from obscured AGN,
as discussed by \citet{May2008}. 
This problem refers to the cumulative contribution to the background of hard 
(energy $\ge$ 3 keV) emission from discrete unresolved sources (AGN), each one emitting 
below the detection threshold of the observation.
 More precisely, for the total of $\sim$\,1450 background
X-ray sources in the Cygnus OB2 survey (Kashyap et al., this issue), we 
found the extragalactic contribution consistent with an N$_{H}$ absorption
$\approx$ 2.0($\pm$0.09)\,$\times$10$^{22}$ cm$^{-2}$, and a power law 
$\Gamma$ index of 1.3($\pm$0.05)  (Flaccomio et al. this issue), which
agree with typical HR $\le$ -0.1 values for AGN.

 In summary, hard unresolved AGN do not play 
a major role in the typical HR of the local diffuse emission, and its hard contribution
appears homogeneous across the field of the observation (see HR map of the obsID 10973). 
This last conclusion agrees with the detailed analysis of the absorption along the line of sight 
of this region, that was independently computed and extensively discussed 
for the foreground, member and extragalactic 
X-ray source populations (Flaccomio et al., this issue).

\subsection{Dust scattering from Cyg\,X-3 emission}
\label{cygx3}

Cygnus~X-3 is known to be a $\gamma$-ray source located in the 
background at a distance of 7.4\,$\pm$1.1 kpc \citep{McCollough2016}, 
more than five times  farther away than the Cygnus OB2 stellar association itself.
Its radiation in X-rays is essentially hard ($\ge$ 3.5 keV; \citealt{Koljonen2010}), although in the 
soft (0.5-2.0 keV) X-ray band it is intense enough that it could mask any prominent 
diffuse structure in the vicinity of the line 
of sight, even when the source PSF has been masked at radii 1.5 times larger than the 99\% Encircled Energy Fraction (EEF).  
We have attempted to mitigate the influence of Cyg~X-3 by disentangling hard and soft X-ray emission 
through the HR coded image, allowing us to reduce the impact of scattered photons 
as well as to explore how far the scattering extends. 

In six of the 40 X-ray observations of the survey, the scattering halo around Cyg X-3 was observed, either 
totally or partially, namely in ObsIds 10939, 10940, 10964, 10969, 10970 and 12099.
In Figure\,\ref{radprof} we show the HR radial profiles and the smoothed images. 
As scattered radiation is dominated by hard photons, even harder than expected from the
unresolved AGN background radiation, the peak of the HR reaches very negative values at the 
centre, with HR$\approx$ -0.8. 
However, as the intensity decays with the inverse square of the radial distance (r$^{-2}$),
at larger distances the influence of scattered X-ray photons from Cyg X-3 decreases  
to the typical HR of the background (HR$_{\rm bkg}$).  At such a distance the radial profile becomes flat.
On average, for all the observations we estimated that the HR$_{\rm bkg}$ levels off at large radial distances 
from Cyg~X-3 at -0.14$\pm$0.07, which is consistent with the value obtained in Section\,\ref{hrsection}.
The typical scattering halo of Cyg X-3 extends to a radius of about $\sim$ 8 arcmin.

We note the existence of some softer sub-structures (HR $\ge$ 0.05) surrounding Cyg X-3, which are 
probably due to scattering 
from dust clouds along the line-of-sight to Cygnus X-3 \citep{McCollough2013}. 
Otherwise, HR values below $-0.1$ cannot realistically be considered part of the
diffuse X-ray emission, as temperatures are required to be above 7 keV, 
which is unconstrained for the limited energy range of the observation.  
In any case, besides the poor photon statistics in such  bands, 
HR values of $\approx -0.1$, or harder, would be better described by a 
simple power-law emission model with spectral index $\Gamma = 1.7$ or lower
 \citep{Corral2011}. 
 Hereafter, conservatively, reliable diffuse X-ray emission 
patterns in the HR maps would be considered  
for those HR values larger than $\approx$ $-0.1$.\\

\section{Large-scale diffuse X-ray map}
\label{mosaic}

\begin{figure*}[ht!]
\centering
\includegraphics[width=8.9cm,angle=0]{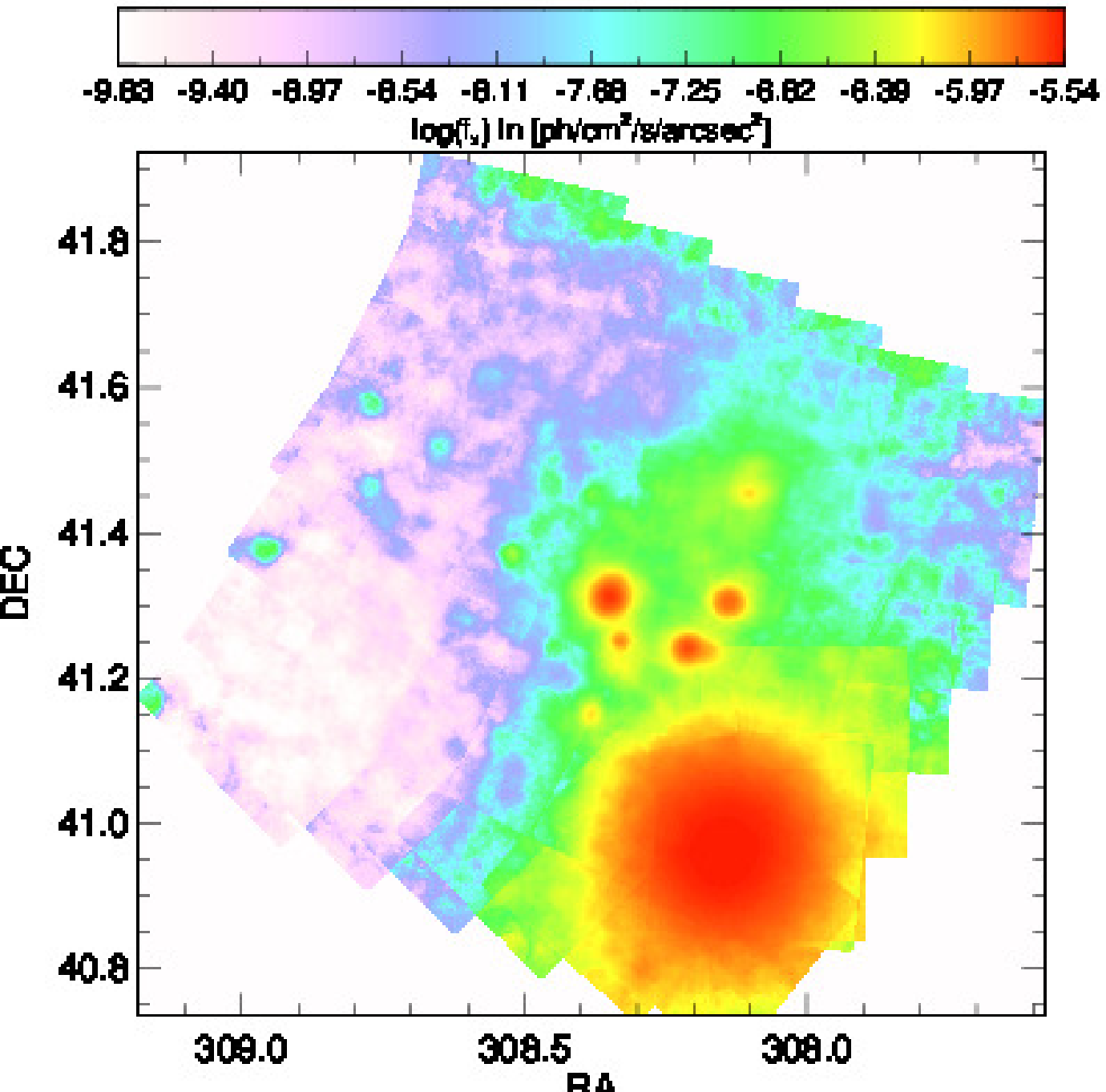}
\includegraphics[width=8.9cm,angle=0]{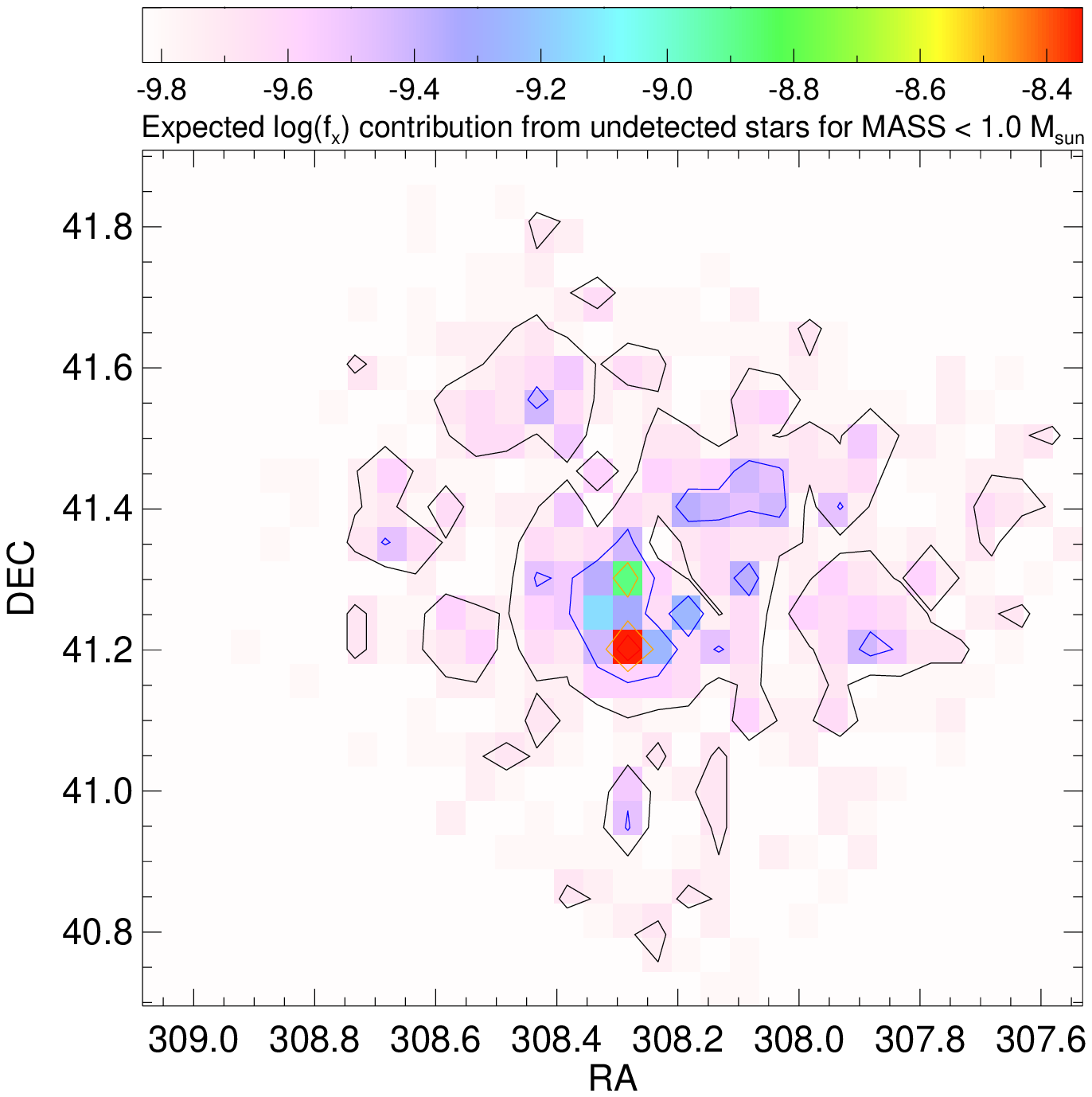}
\caption{\small Left: The X-ray diffuse emission map for the full Cyg OB2 FOV in the [0.5-7.0]~keV band.
Right: Contours in $\log({\rm f_{unresolved}})$ indicating the intensity levels 
expected from the X-ray flux of unresolved stars.}
\label{fxunr}
\end{figure*}

The morphology and energetic  large-scale appearance of diffuse 
 X-ray emission in the Cygnus~OB2 association is shown in Figure\,\ref{fxunr}-left. 
It is a mosaicked, point source-removed image in the [0.5-7.0] keV energy range.  
Similar images were also made 
in three discrete energy bands: Soft [0.5-1.2] keV, Medium [1.2-2.5] keV and Hard [2.5-7.0] keV. 
These images include the complete set of ACIS-I observations of our survey 
and cover an area of about 1~deg$^2$ (see Section\,\ref{discussion}).

Mosaicking was performed using the Montage software to generate a list of images 
with corresponding WCS information. 
Before reprojection, we computed the individual emission levels required to 
match overlapping images. To do this, we determined the average emission level of two or multiple
overlapped observations and scaled each image to match this average emission level.
In cases of multiple overlapped images, the average diffuse emission level 
was computed by equal weighting averages.
While more complex overlapping functions (e.g., cubic splines) might 
be employed, and can in principle work well to match 
background levels in overlapping observations with X-ray point 
sources, we found that spline smoothing functions at the faint X-ray diffuse emission levels that characterize 
the Cygnus~OB2 observations can be numerically ill-conditioned, producing biased spline coefficients, 
not only resulting in mismatched scaling at the borders of each observation, but also 
producing fake large scale structures.

In the process of obtaining a full coverage diffuse X-ray emission map for the 
observed Cygnus OB2 region, and accounting for all the issues that could affect the diffuse X-ray emission level, 
we still need to account for one other possible underlying contamination problem, which is the contribution from the 
unresolved population of low mass stars, in the stellar mass regime in which our survey detection is incomplete.

\subsection{Unresolved stellar population vs. diffuse X-ray emission}

One of the major problems facing the detection of diffuse
X-ray emission in SFRs is the underlying contribution
from unresolved sources that individually give rise to
counts that fall below the detection threshold of the region. 
Three different contributions are addressed separately:\\ 

\noindent
i- \textit{Background contribution:}\\
Guarcello et al.\ and Wright et al. (this issue) found that about 78\,\%\ 
(6149 / 7924) of X-ray sources detected in the {\it Chandra} survey were classified as Cygnus\,OB2 members. 
For background objects (1304 sources classified), the contribution 
from undetected AGNs remains under the detection threshold all over the entire set of observations, 
as was computed at the centre of our Cygnus~OB2 region using a deeper (100 ksec) observation \citep{Albacete2007}.\\

\noindent
ii- \textit{Foreground contribution:}\\
The case of undetected foreground stars is more controversial, because they appears softer than that 
of the background sources (see Flaccomio et al., this issue), and so it is expected to be more difficult 
to distinguish from the spectrum of the diffuse emission. 
 Based on the classification analysis of Chandra X-ray sources in the Cygnus OB2 region, 
we have identified $\sim$471 X-ray sources ($\sim$\,6\% of the total) as foreground stars
(Kashyap et al., this issue).
The stacked spectrum of all individually-detected foreground stars towards the Cygnus OB2 region shows a 
soft thermal plasma emission with typical temperature and absorption of kT = 0.77 keV and
N$_{\rm H}^{\rm upper \,limit}$$\approx$0.2$\times$10$^{22}$ cm$^{-2}$, respectively (see Flaccomio et al., this issue).
The total unresolved foreground stars are likely to have a similar spectral shape that could easily 
contribute to the observed soft diffuse X-ray emission. However, the clear anti-correlation between 
diffuse X-ray emission and observed colder gas-dust structures in the Cygnus OB2 association (see discussion in 
Section\,\ref{discussion}), suggests that the soft diffuse emission we detected is mainly associated to the Cygnus OB2 region. 
Alternatively, we used the investigation of \cite{Getman2011} that estimates the star foreground
unresolved population of 200,000 toward Carina (2.3 kpc and Area$\sim$ 1.42 sq.deg.). 
 By scaling to the Cygnus OB2 region (1.4 kpc and Area\,$\sim$\,0.97 sq.deg.) 
we estimate $\sim$\,18,500 unresolved foreground stars that would be in the projected area of the survey. 
They would emit at levels below the typical detection threshold of the survey of 6$\times$10$^{-5}$ cnts/s ($\sim$ 3 photons) 
\citep{Wright2015}. So, the total number of expected photons from unresolved foreground stars is less than or equal to 
3$\times$18,500 $\approx$ 56,000 photons, or just $\sim$ 8\% of the total of counts ($\sim$ 710,000) 
observed in the entire 0.97 sq.deg Cygnus OB2 diffuse X-ray emission map.\\

\noindent
iii- \textit{Stellar Cygnus OB2 contribution:}\\
In order to compute the contribution to the observed X-ray diffuse emission level from unresolved low-mass stars 
belonging to the Cygnus OB2 region, we adopted the completeness limit for the survey from \cite{Wright2014a}.
With some small spatial variations (generally 10\% or less in terms of flux), 
the X-ray luminosity completeness limit for our Cygnus OB2 survey area
ranges from 50\% at 1.4$\times$10$^{30}$ ergs s$^{-1}$ to 90\% at 4$\times$10$^{30}$ ergs s$^{-1}$.
The actual percentage of stars detected is different for stars of different masses because the 
X-ray luminosity distribution is mass-dependent.
For the entire survey area the completeness is 50\%\ at 0.6 M$_\odot$ and 90\% at 1.3 M$_\odot$, respectively \citep{Wright2014a}.
For our purposes, the X-ray luminosities of stars were computed by adopting the X-ray conversion factor 
(CF), i.e. the ratio between unabsorbed flux (f$_{\rm ua}$) in ergs to the absorbed flux in photons 
(f$_{\rm abs}$ [ph]), that corresponds to a value of $5.4\times 10^{-9}$~[erg/ph] (see Flaccomio et al, this issue).

We assumed the X-ray luminosities of the stellar population of Cygnus OB2 stars at masses 
in which our survey is essentially complete are the same as stars of the same mass in the Orion Nebular Cluster (ONC).  
We then used the results of the Chandra Orion Ultradeep Project (COUP), that is essentially complete at all 
masses above 0.3 M$_\odot$ \citep{Feigelson2005, Getman2005, Preibisch2005}, to infer the 
signal from stars in Cygnus~OB2 with masses below 1.0 M$_\odot$ for which we are complete to a level of about 85\%.

The expected X-ray luminosity contribution for stars with masses in the range [0.3\,--\,1.0] M$_\odot$, 
per star in the ONC is L$_{\rm uc}^{\rm ONC}$ $\sim$ 5.8$\times$10$^{30}$ erg\,s$^{-1}$. 
At the distance of Orion (d $\sim$ 450 pc), and accounting for our CF [erg/ph], we obtained the 
expected flux per detected star, f$_{\rm uc}^{\rm ONC}$=4.5$\times$10$^{-5}$ [ph/s/cm$^2$/star].
For COUP stars with a visual extinction A$_{\rm v}\le$ 5 mag and masses below  1.0 M$_\odot$, 
we get a total population of 84 stars. Applying the same restriction for the Cygnus OB2 members, 
we find 786 stellar members with masses in the range [0.3\,--\,1.0] M$_\odot$.
This implies that Cygnus OB2 is $\sim$9.2 times more massive than Orion for the same mass range. 
By adopting a distance of 1450~pc to Cygnus OB2, the total X-ray contribution for unresolved
stars in the range  [0.3--1.0] ~M$_\odot$ and over the entire observed area ($\sim 1 \rm deg^2$) is 
f$_{\rm uc}^{\rm CygOB2}$ $\sim$4.6$\times$10$^{-6}$ [ph/s/cm$^2$/star].
In the left panel of Figure\,\ref{fxunr}, we illustrate the Cygnus OB2 source density map binned at 0.05 deg$^2$, 
weighted by the expected X-ray flux contribution of undetected stars. 
Thus the expected flux, f$_{\rm uc}^{\rm CygOB2}$, should be divided by the unit of area adopted for binning,
dens\_area= (0.05$\times$3600)$^2$ [stars/arcsec$^2$].
Finally, for each detected low-mass star the expected X-ray flux emission from unresolved
stars per arcsec$^2$ (f$_x^{uc}$) 
is f$_{\rm uc}^{\rm CygOB2}$/dens\_area = 1.4$\times$10$^{-10}$ [ph/s/cm$^2$/arcsec$^2$],
i.e.\ (log(f$_x^{UC}$) = -9.84). Note the peak of the expected X-ray flux contribution is
log(max(f$_x^{UC}$))= -8.34 [ph/s/cm$^2$/arcsec$^2$] and this occurs only in a single region of 0.05 deg$^2$ centered 
at RA=308.3~deg and DEC=41.2~deg (Figure~\ref{fxunr}-right). 
However, this value is overestimated because we
made use of a completeness function computed from 120~ks simulations \citep{Wright2015}, while at this
central position our survey has a nominal summed exposure of $\sim$ 220 ksec, 
leading into a deeper source detection threshold. 

With this estimation, and at the same time, by comparison with our X-ray diffuse emission map, 
computed using the same spatial binning of 0.05~deg$^2$
(see figure\,\ref{fxunr}-right), the detected X-ray diffuse emission patterns in the survey are 
above log(f$_{\rm x}$)$\approx$-8.2. Thus the contribution
from the unresolved stellar population, in the worst case,  remains 
a factor 2.5 fainter than the observed emission.  For the vast majority of the survey 
area, it is at least an order of magnitude lower, 
confirming detected X-ray diffuse emission structures are real, and not
affected by the contribution of X-ray emission from undetected  low-mass stars of the region.\\

\section{Spectral analysis of diffuse X-ray emission} 
\label{spectral}

\begin{figure}[!ht]
\includegraphics[width=8.6cm,angle=0]{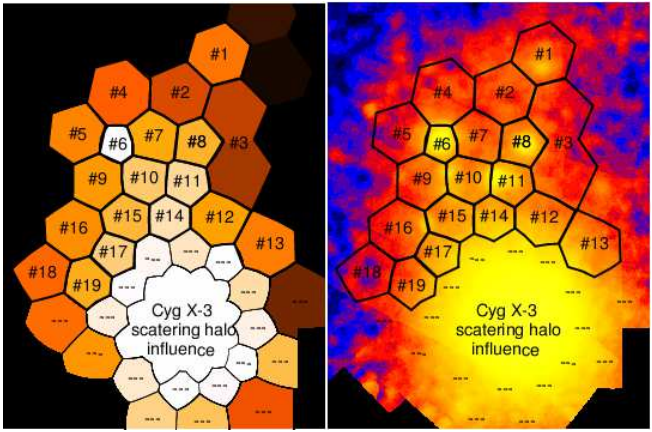}
\caption{\small Left: Tessellated X-ray surface brightness of the Cygnus OB2 region. 
The achieved S/N is 60 and regions were computed to perform spatially-resolved 
X-ray spectral fitting. Regions labeled with dashed lines were not taken into account 
due to the influence of the background scattered X-ray radiation from Cyg X-3
(see discussion in section\,\ref{cygx3}). 
Right: Diffuse X-ray mosaic events used for spectral extraction.} 
\label{wvtimage}
\end{figure}

X-ray spectral fitting of diffuse emission is a difficult undertaking, and 
requires sufficient X-ray photons to provide meaningful constraints
 on the X-ray spectral model parameters. The most straightforward procedure is to increase the 
source extraction areas to gain larger signal-to-noise ratios.  To this end, 
we tessellated the whole image mosaic in the [0.5-7.0] keV band to achieve 
surface brightness regions with S/N ratio greater than 60. 
Figure~\ref{wvtimage} shows the tessellate-generated image made using the Weighted Voronoi Tessellation (WVT)
Binning code  \citep{Diehl2006}.
These tessellated regions can be readily translated 
into region files and used to extract events within CIAO.
In order to avoid the contamination produced by the bright background X-ray scattered halo 
of Cyg\,X-3, all tessellated regions 
within a 8~arcmin radius of its position (RA $\approx$ 20:32' , DEC $\approx$ +40:57')
 were excluded from the spectral analysis.
The choice of this distance was discussed earlier in Section\,\ref{cygx3}. 
However, for some tessellated regions, even beyond 8 arcmin, the fractional 
X-ray contribution would produce marginal contamination in the 
diffuse X-ray spectra. We account for this issue 
in the spectral model fitting and subtract residual Cyg\,X-3 signal from the total [0.5-7.0] keV diffuse
emission. The spatially resolved spectral fitting models and parameters are
presented in Table\,1.

\begin{figure*}[!ht]
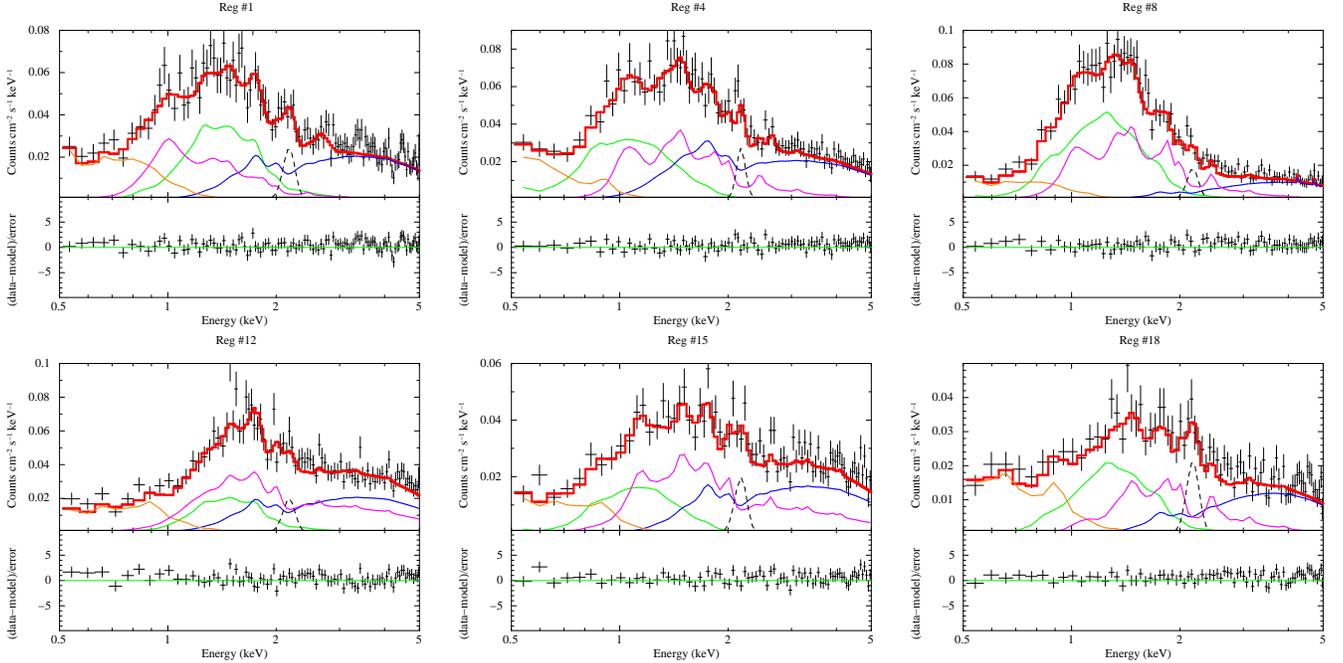

\centering
\includegraphics[width=4.4cm,angle=270]{fig12_up_left.ps}
\includegraphics[width=4.4cm,angle=270]{fig12_up_cent.ps}
\includegraphics[width=4.4cm,angle=270]{fig12_up_right.ps}\\
\includegraphics[width=4.4cm,angle=270]{fig12_down_left.ps}
\includegraphics[width=4.4cm,angle=270]{fig12_down_cent.ps}
\includegraphics[width=4.4cm,angle=270]{fig12_down_right.ps}
\caption{\small Sample of some tessellate spectra for regions- \#1, 4, 8, 12, 15 and 18. 
Models that describe diffuse X-ray emission are indicated with 
red solid lines. The thinner colored lines represent the decomposition of the total 
emission into different models: Super-Soft (SS - orange), Soft (S - green),
Moderate (M - Magenta) and Hard (H - blue).
Note: Black dashed line correspond to the residual instrumental line at $\sim$2.1 keV 
from background stowed spectra, which has not be taken into account for the 
event/stowed-bkg normalization (see\,\ref{bkg} for discussion).} 
\label{xspecplots}
\end{figure*}

The extraction of the X-ray spectrum for each tessellated region 
was achieved by using the {\rm specextract} CIAO task from the respective
diffuse event files. All spectra were properly weighted by the appropriate calibration files 
(the so-called ARFs and RMFs), that account for the many partially-overlapping ObsIDs.
Background X-ray spectra were computed by using "stowed calibration events files".
All X-ray spectra were grouped to reach a minimum Signal-to-Noise Ratio (SNR) per bin of 1, 
which produce unbiased best-fit values for the fitting procedure \citep{Albacete-Colombo2016}.

For spectral fits, we used a suite of XSPEC spectral models \citep{Arnaud1996}
to adequately account for possible combinations of emission components 
affected by equivalent hydrogen NH absorption column.
The latter was included using the {\sc TBABS} (Tuebingen-Boulder - TB) model \citep{Wilms2000}, 
that is composed by a combination of N(HI) (atomic hydrogen) and 2\,N(H2) (molecular hydrogen). 
We tested the differences in the use of thermal emission models such as {\sc "PSHOCK"} (PS), 
that is an averaged-abundance plane-parallel shock 
in non-collisional ionization equilibrium (NEI), and combination of {\sc "APEC"} (AP)
collisional ionization equilibrium (CIE) plasma models \citep{Smith2001}.
 
\begin{table*}[!ht]
\label{spectral_tab}
\caption{Spectral model parameters of diffuse X-ray emission}
\begin{center}
\begin{tabular}{lllllllll}
\hline \hline
Reg.	 & 	Model &	N$_{\rm H}$ [$\times$10$^{22}$ cm$^{-2}$] &kT [keV]  & Norm.&  Z & 
Flux ($\times10^{-12}$ cgs)&L$_{\rm x}$ ($\times$10$^{33}$ cgs) \\
\#	&	  &Diffuse / Bkg.   &Diffuse / Bkg.	&  [cm$^{-3}$]& [Z$_\odot$]	&	
Diffuse / Bkg. & Diffuse / Bkg.  \\			

\hline
1 	&3T / bkg		&0.34$-$0.58$-$1.37$ / $2.6 & 0.11$-$1.01$-$0.48$ / $44.5 & 0.26 &1.3 		& 6.07 / 3.13 & 1.52  / 0.78 \\
2 	&3T / bkg		&0.50$-$0.59$-$1.53$ / $2.6 & 0.10$-$0.76$-$0.42$ / $34.4 & 0.40 &1.7 		& 10.9 / 2.99 & 2.74  / 0.75 \\
3 	&3T / bkg		&0.46$-$1.05$-$1.67$ / $5.8 & 0.11$-$1.24$-$0.50$ / $[64.0]& 0.56 &3.4 		& 29.2 / 9.12 & 7.34  / 2.29 \\
4 	&3T / bkg		&0.33$-$0.50$-$1.35$ / $1.6 & 0.09$-$0.31$-$1.23$ / $[64.0]& 0.11 &1.7 		& 5.80 / 2.75 & 1.46  / 0.69 \\
5 	&3T / bkg		&0.44$-$1.04$-$1.48$ / $5.6 & 0.12$-$1.22$-$0.47$ / $56.1  & 0.23 &3.7 		& 11.79 / 3.01& 2.96  / 0.75 \\
6 	&3T / bkg		&0.57$-$1.00$-$1.05$ / $[10] & 0.13$-$0.56$-$1.21$ / $13.7   & 0.13 &2.1 		& 6.77 / 1.76 & 1.71  / 0.44 \\
7 	&3T / bkg		&0.18$-$0.66$-$0.98$ / $2.7 & 0.08$-$0.91$-$0.42$ / $[64.0]& 0.23 &1.4 		& 3.34 / 2.13 & 0.85  / 0.53 \\
8 	&3T / bkg		&0.33$-$0.80$-$1.39$ / $4.6 & 0.11$-$0.38$-$1.02$ / $57.3  & 0.12 &2.1 		& 7.25 / 1.86 & 1.82  / 0.46 \\
9 	&3T / bkg		&0.29$-$0.76$-$1.44$ / $2.8 & 0.11$-$0.39$-$1.08$ / $[60]   & 0.11 &2.1 		& 4.26 / 2.32 & 1.07  / 0.58 \\
10 	&3T / bkg		&0.45$-$1.23$-$1.38$ / $4.5 & 0.13$-$1.21$-$0.46$ / $43.3  & 0.14 &1.6 		& 7.44 / 2.74 & 1.87  / 0.68 \\
11 	&3T / bkg		&0.44$-$1.00$-$1.36$ / $4.2 & 0.10$-$1.11$-$0.51$ / $42.7  & 0.44 &1.9 		& 6.69 / 2.37 & 1.68  / 0.59 \\
12$\dag$	&2T\,/\,Cyg\,X-3+bkg  &0.49$-$0.97$ / $1.7$-$2.7 & 0.12$-$0.45 $ / $11.4$-$[64] & 0.25 &3.8 	& 9.31 / 3.15 & 2.34  / 0.79 \\
13$\dag$	&2T\,/\,Cyg\,X-3+bkg &0.40$-$0.71$ / $2.9$-$2.6 & 0.11$-$0.53$ / $10.8$-$[64] & 0.26 &1.8 	& 9.30 / 2.87 & 2.33  / 0.72 \\
14$\dag$	&2T\,/\,Cyg\,X-3+bkg &0.52$-$1.60$ / $2.9  	& 0.10$-$0.41$ / $61.8 & 0.27 &1.1 			& 9.96 / 3.16 & 2.50  / 0.79 \\
15 	&3T / bkg		&0.43$-$0.74$-$0.83$ / $2.6 & 0.10$-$0.27$-$5.7$ / $39.8  & 0.26 &3.8 		& 5.25 / 2.56 & 1.32  / 0.64 \\
16 	&3T / bkg		&0.38$-$0.80$-$1.81$ / $3.4 & 0.11$-$0.38$-$0.68$ / $7.8  & 0.16 &1.8 		& 5.97 / 5.02 & 1.51  / 1.26 \\
17$\dag$	&2T\,/\,Cyg\,X-3+bkg&0.50$-$1.86$ / $2.6  	& 0.11$-$0.51$ / $10.8 & 0.50 &1.7 			& 5.26 / 2.90 & 1.33  / 0.72 \\
18 	&3T / bkg	&0.34$-$0.98$-$2.38$ / $4.1 & 0.13$-$0.44$-$1.21$ / $27.4  & 0.04 &1.7 	& 5.46 / 2.38 & 1.37  / 0.59 \\
19$\dag$	&3T\,/\,Cyg\,X-3+bkg &0.38$-$1.26$-$1.61$ / $2.6  & 0.11$-$0.32$-$1.26$ / $[64] & 0.29 &3.9 & 14.6 / 2.66 & 3.67  / 0.66 \\
\hline
\hline
Stacked &	 Model	&  N$_{\rm H}$-components & kT-components    & Norm. & Z		 	& Flux  &  	 L$_{\rm x}$ & $\%$\,Diff.  \\	
\hline
SS	     & Super-Soft &  0.43 $\pm$ 0.09   & 0.11 $\pm$ 0.01 &  2.53&      1.3	    		        &  60.4 $\pm$ 7.5  &          15.2\,$\pm$\,1.9	& 37\,$\pm$\,3    \\  
S	     &            Soft &  0.80 $\pm$ 0.22   & 0.40 $\pm$ 0.08 &  1.79&     1.8    			&  75.1 $\pm$ 10.1  &          18.9\,$\pm$\,2.7  	&  46\,$\pm$\,5  \\  
M	     & Moderate   &  1.39 $\pm$ 0.28   & 1.18 $\pm$ 0.18 &   0.43&     3.1      			&   31.0 $\pm$ 5.1 &            7.8\,$\pm$\,1.3  	&  17\,$\pm$\,2 \\  
H	     & Hard          &  2.70 $\pm$ 0.42   & 45 $\pm$ 19         &   0.09&      1.0          		&   56.0 $\pm$ 16.7 &          14.1\,$\pm$\,4.2 	&   $--$       	\\  
Diffuse   &	SS+S+M	&                                & 				& 4.76 &   [1.2]    			& 166.7 $\pm$ 26.1 &  	41.95\,$\pm$\,6.6 & 100 	\\	
\hline
\end{tabular}

The "bkg"---the hardest component of the diffuse emission spectra---refers to the contribution 
from unresolved AGN background emission. A thermal model approximation is sufficient  
to describe the hard emission (see text for discussion). The column 5 (Norm.) and 6 (Z) refers to the
normalization parameter and the abundance of the model, respectively.
The presence of diffuse non-thermal emission is addressed in Section\,\ref{discussion}.
Both flux and luminosity were computed for the 0.5 - 7.0 keV energy range. $\dag$ refers to 
spectra affected by Cyg X-3 scattered photons, which were also modelled by a thermal component.
 Last column of the bottom table indicates the percentage of SS, S and M contribution to the 
total diffuse X-ray emission.
\end{center}
\end{table*}

Unfortunately, the statistics of the spectra impose a limitation on 
the number of spectral models and free parameters 
that can be usefully constrained.  We initially used the PS model to fit the 
softer ("Super-Soft" - SS) component (kT1), that gives short ionization timescales 
(from $\tau_{\rm u}$ $\sim$ 10$^{-10}$ to 10$^{-11}$ s\,cm$^{-3}$), 
implying a low-density highly non-equilibrum plasma (NEI). 
This interpretation supports intense and recent stellar winds and ISM shock 
interactions \citep{Smith2010a}. Otherwise, 
"Soft" (S) and "Moderate" (M) energies were successfully described by two AP 
plasma models (kT2 and kT3) that adequately fit the observations for intermediate temperatures 
of each tessellated region. Finally, a fourth "Hard" (H) CIE plasma model (kT4) accounts for 
background non-resolved AGN---faint and hard---X-ray emission.  
Such a hard thermal component was able to match most of the hard diffuse emission, 
even though AGN are expected to be dominated by non-thermal emission and are usually well modelled with power-law
spectral shapes. However, the use of a thermal plasma spectral shape (kT4) may also be
fitting more than just the unresolved AGN emission \citep{Townsley2011a}.
Several attempts with simple (1T or 2T) models do not adequately fit the shape of the diffuse X-ray spectra, so
we adopt a 4T combined model for spectral fitting, that is written with an XSPEC expression:\\
\begin{equation}
TB_1 \times PS_1 + TB_2 \times AP_2 + TB_3 \times AP_3 + TB_4 \times AP_4
\end{equation}

Models with variable abundance, e.g. {\sc "VPSHOCK"} and {\sc "VAPEC"}, were not used
as the number of free channels in the spectra is insufficient to discriminate adequately for  
differential abundance contributions from single elements. Otherwise, the fixed solar metal abundance pattern
with scalable metallicity, Z, was not allowed to go below solar (Z$=1$) as it is unlikely that massive
star winds could produce sub-solar plasmas abundances \citep{Strickland1998, Pittard2010}.
In fact, sometimes we slightly improved the goodness of the fit by allowing super-solar abundances, 
even if the actual abundance values were often not well-constrained, and/or eventually fixed.
The spectral analysis was performed in an interactive way, 
with metallicity thawed, but carefully fitting in the restricted range of Z$=1.0$--5.0.

For some tessellated regions (\# 12, 13, 14, 17 and 19, see Table\,1), 
the fourth hard (kT4) component helps us to disentangle the marginal contribution of the 
diffuse background ($\sim$ 9 kpc) scattering emission from Cyg~X-3.
In these cases, the absorption associated with the hard thermal component would 
be representative of a large total N$_{\rm H}$ absorption column, so
it was initially set at 1.8$\times$10$^{22}$ cm$^{-2}$, but restricted in the fit 
to the range 0.8--10.0$\times$10$^{22}$ cm$^{-2}$.

The goodness-of-fit ($\chi^2$/dof)  obtained is generally acceptable 
(ranging from $\sim$1.0--1.1) for most of the tessellated regions. 
However, some cases of $\chi^2$/dof $\sim$1.2--1.3 are probably associated 
with an ill-constrained emission model for scattered hard X-ray photons 
from the Cyg X-3 background radiation. 

In Figure\,\ref{xspecplots}, we show of six example X-ray diffuse spectra
that exhibit different spectral characteristics.
In Table\,1, we give the best-fit spectral parameters for the diffuse X-ray emission tessellated areas.
Not surprisingly, three independent absorption models shape the apparent diffuse X-ray brightness of the region.
 For the stacked spectrum we get absorption values N$_{\rm H}^1$, N$_{\rm H}^2$, and 
N$_{\rm H}^3$ of 0.43 (1$\sigma$=0.09), 0.80 (1$\sigma$=0.22), 
1.39 (1$\sigma$=0.28) $\times$10$^{22}$ cm$^{-2}$, respectively, and temperatures of the components kT$_1$, kT$_2$ and 
kT$_3$ of 0.11(1$\sigma$=0.01), 0.42(1$\sigma$=0.08) 
and 1.21(1$\sigma$=0.18), respectively. This is direct evidence of the existence of a wide range of multi-temperature gas in the region.
Otherwise, the hardest component is adequately described by a typical N$_{\rm H}^{4}$\,$\approx$2.7$\times$10$^{22}$ cm$^{-2}$
and a hard contribution from the unresolved AGN background population that can be represented by a $\sim$ 2.8 to 15~keV plasma.

\begin{figure}[!h]
\centering
\includegraphics[width=6.5cm,angle=270]{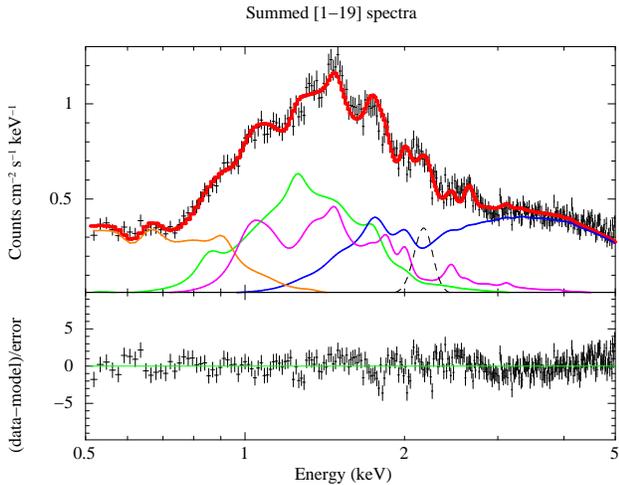}
\caption{\small Stacked diffuse X-ray spectrum for the Cygnus OB2 region.  The total (0.5\,--\,7.0 keV) 
 diffuse X-ray fitted model (thick$-$red) has an intrinsic L$_{\rm tot}$ of 5.6$\times$10$^{34}$ erg s$^{-1}$. 
The total emission is decomposed into different models: Super-Soft (SS - in orange), Soft (S - green),
 Moderate (M - Magenta) and Hard (H - blue).
 } 
\label{true_spectra}
\end{figure}

Spectral analysis of the combined (``stacked'') spectrum of the full region was 
performed using a four-temperature thermal input model. 
Results are consistent with those obtained from the averages computed for spatially-resolved
spectra. The total intrinsic diffuse X-ray luminosity is L$_{\rm x}$=5.6$\times$10$^{34}$ 
erg\,s$^{-1}$ for the 0.5\,--\,7.0 keV energy range. 
As shown in Figure\,\ref{true_spectra}, the total diffuse spectrum is successfully reproduced by a mix of 
the Super-Soft (SS)+ Soft (S) + Moderate (M) + Hard (H) plasma emission models.  Unfortunately, 
there are also some regions of SS emission away from the 
centre of the Cygnus OB2 association that are too faint to construct 
tessellated regions for spectral extraction at S/N$>$1. 

In contrast with the first three thermal components (SS, S and M), the H model appears highly absorbed 
($\sim$ 2.6$\times$10$^{22}$ cm$^{-2}$) and extremely hard (kT$>$ 15 keV), as expected 
from the unresolved AGN background population and/or---for some cases---the 
scattered radiation from Cyg\,X-3. This hard component is then not part of the local diffuse 
X-ray emission of the region, so we have disentangled such component to estimate a background X-ray 
luminosity of L$_{\rm x}^{\rm H}$ ($\sim$ 1.41$\times$10$^{34}$ erg\,s$^{-1}$).
Certainly, hard X-rays might originate in the Cyg~OB2 association. In section\,\ref{NT} we discuss 
plausible non-thermal emission mechanisms present in the local diffuse X-ray emission of the region. 

We consider the truly diffuse X-ray emission of the Cygnus OB2 region to be composed
of the contribution from SS + S + M models, which have a total combined X-ray luminosity 
of L$_{\rm x}^{\rm diff}$=4.2$\times$10$^{34}$ erg\,s$^{-1}$ for the 0.5-7.0 keV energy range.
The softer SS component is compatible with a temperature of 
0.11 keV and an low absorption column of 0.42$\times$10$^{22}$ cm$^{-2}$, and which appears spatially
related to the shock interaction between winds from massive stars and the local 
ISM (see section\,\ref{discussion}).
Its relative contribution to the total X-ray diffuse emission of the region is 
$\sim$37\% (L$_{\rm x}^{\rm SS}$ $\sim$ 1.52$\times$10$^{34}$ erg\,s$^{-1}$), although owing to the 
scattered soft emission too faint for quantitative analysis mentioned above, the true 
SS flux could be larger. In fact, SS emission is just observed where absorption is low enough, being
completely absorbed at other parts of the region. The S and M components, with respective temperatures of 0.40 and 1.18 keV, 
appear 2.8 to 3.2 times more absorbed (1.18 and 1.30 $\times$10$^{22}$ cm$^{-2}$, respectively) than the SS emission.
The S component contributes with the 46\% (L$_{\rm x}^{\rm S}$ $\sim$ 1.89$\times$10$^{34}$ 
erg\,s$^{-1}$) and the M component 17\% (L$_{\rm x}^{\rm M}$ $\sim$ 0.69$\times$10$^{34}$ erg\,s$^{-1}$) 
of the total diffuse X-ray luminosity of the region.\\

\section{Discussion}
\label{discussion}

\begin{figure*}[!ht]
\centering
\includegraphics[width=18cm,angle=0]{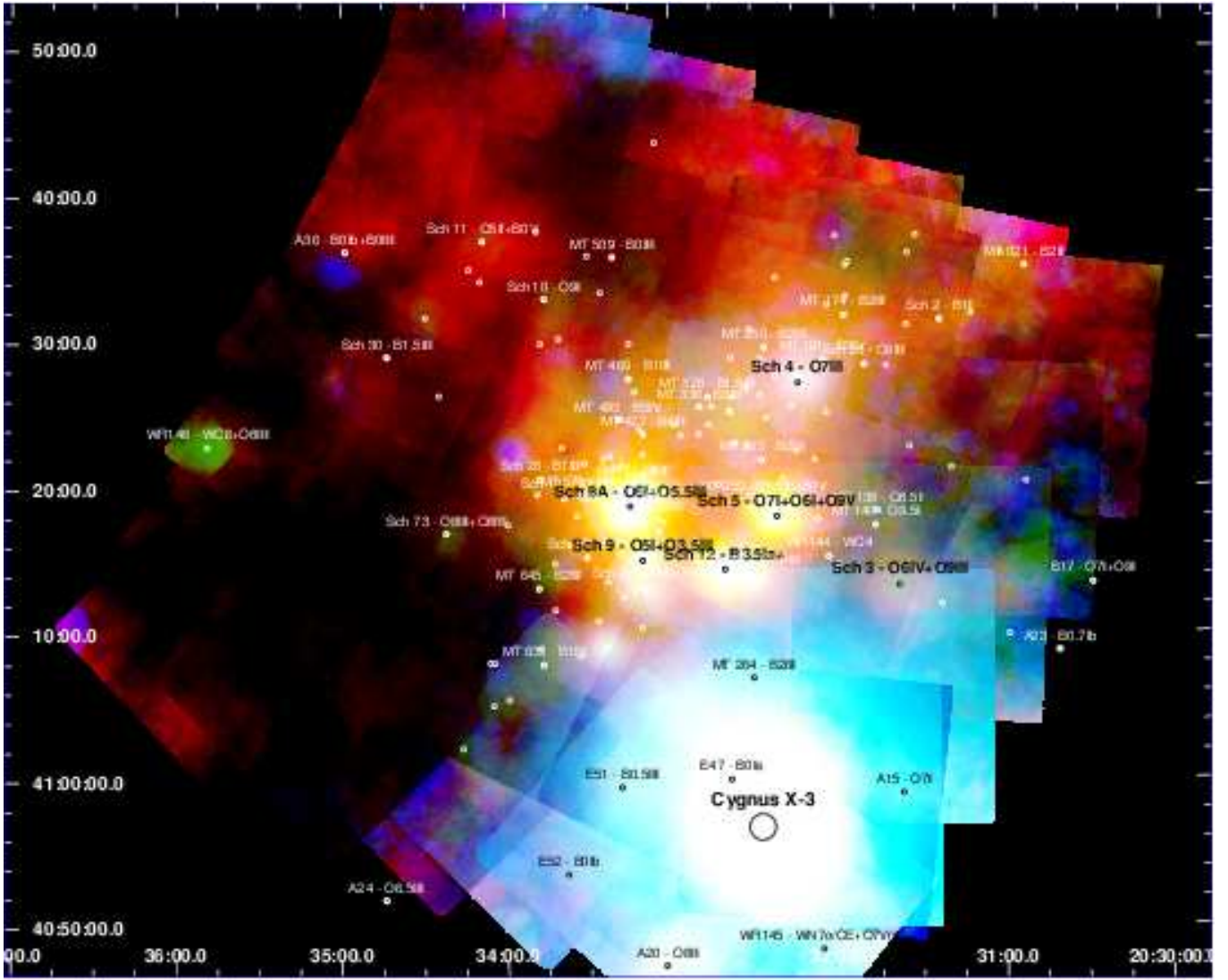}
\caption{\small The Cygnus OB2 diffuse emission map in the 0.5\,--\,7.0 keV band. The full band 
is coded in color so that
soft [0.5\,--\,1.2] keV emission appears in red, and medium [1.2\,--\,2.5] keV in green, while
hard [2.5\,--\,7.0] keV emission appears in blue. 
Small white circles indicate the massive stellar content of the region. Names and labels
indicate the evolved stars with more intense stellar winds.
Names are omitted for main sequence O- and B-type stars that have less massive winds. 
The X-ray mosaic intensity is logarithmically scaled, and the region shown is approximately 
$1.3 \times 1.3$ degrees, with North up and East to the left.}
\label{largescale}
\end{figure*}

This work has demonstrated, for the first time, the existence of true diffuse X-ray emission in the Cygnus~OB2
stellar association. 
It is of interest to examine possible mechanisms responsible for the observed diffuse emission and its
spatial morphology, which is probably related to a mix of different, thermal and/or non-thermal, 
physical mechanisms acting separately, but sharing a single spatial region. 
In Figure~\ref{largescale} we show a global view of the diffuse X-ray emission that is dominated 
by a mix of soft [0.5\,--\,1.2] keV and intermediate hardness [1.2\,--\,2.5] keV X-rays, in agreement with
three thermal contributions that we obtained from spectral analysis.
The 2.5\,--\,7.0 (hard) keV energy band is instead largely dominated by the background contamination 
(see section\,\ref{hrbgk}) and/or, in some regions,  scattered radiation from Cygnus X-3 
(see section\,\ref{cygx3}). 

\subsection{Thermal contribution to diffuse X-ray emission}
\label{thermal}

The diffuse emission closely follows the spatial 
distribution of massive stars \citep{Wright2015a}. The most widely accepted physical 
mechanism for production of diffuse hot gas 
is by multiple interactions of the winds from massive stars with the 
ambient ISM \citep[e.g.,][]{Canto2000}. This kind of ISM--wind interaction occurs under 
the action of stellar wind momentum so as to produce low ISM densities, such that the X-ray diffuse gas
would be characterized by high-pressure (low ISM density) and high temperature.
However, it is remarkable that the Soft [0.5--1.2] keV contribution, which is somewhat fainter than 
intermediate energy band emission, appears much more dispersed and also less 
confined at the locations of evolved massive stars.
This result suggests that the cumulative influence of intense massive stellar winds 
act to fill and heat the surrounding ISM, injecting enough thermal energy to drive 
outward via expanding turbulent diffusive motions of hot gas on scales of several parsecs,
even in places absent of massive stars \citep{Dwarkadas2013}.

\begin{figure*}[!ht]
\centering
\includegraphics[width=18.8cm,angle=0]{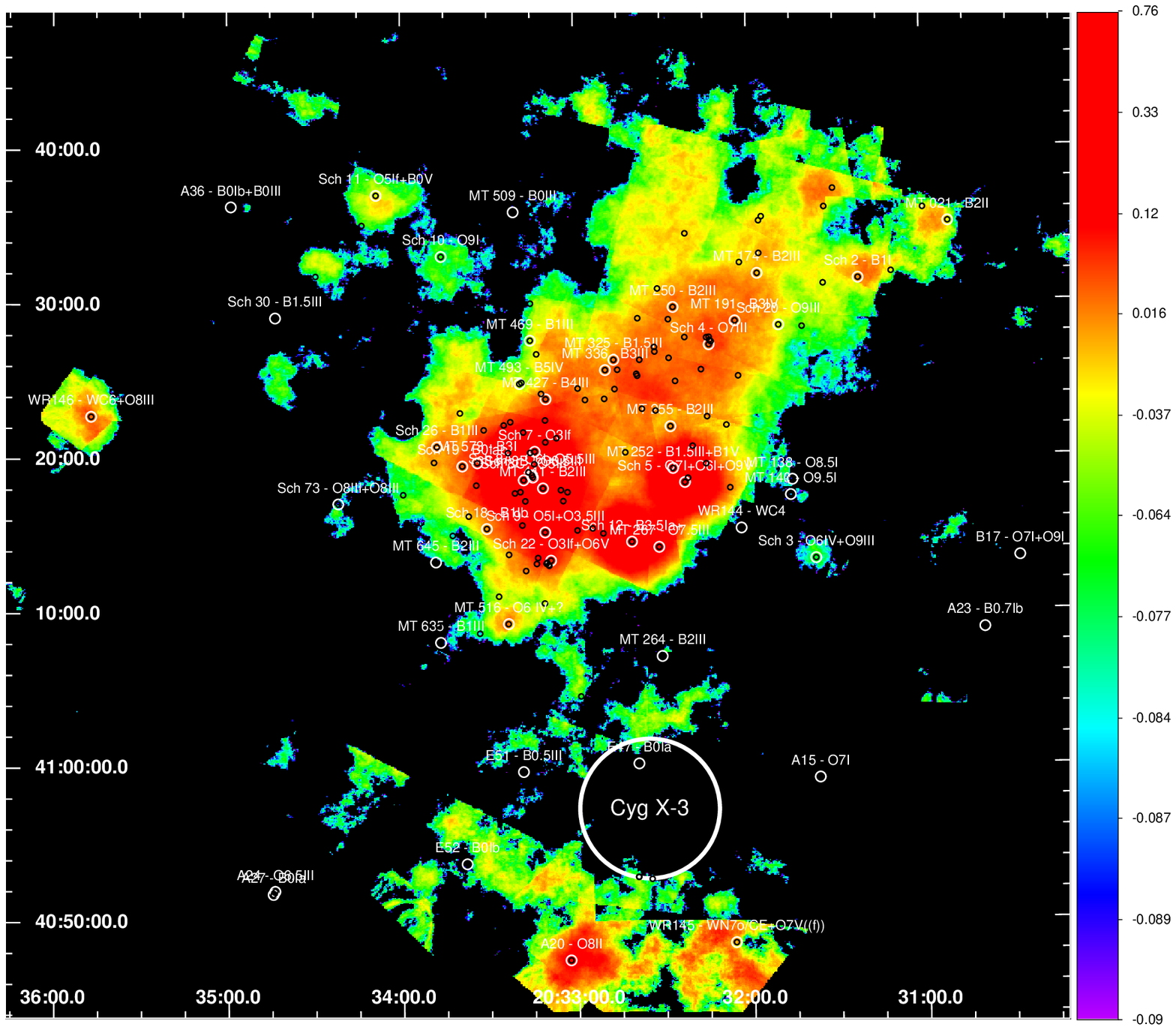}
\caption{\small The Cygnus OB2 hardness ratio (HR) diffuse map in the 
soft [0.5\,--\,2.5] keV and hard [2.5\,--\,7.0] keV bands. Spatial regions of diffuse gas 
that correspond to HR values lower than $- 0.1$ were discarded (see section\,\ref{hrsection}).
Small black circles indicate the MS massive stars of the region. Names and labels
in white indicate the evolved stars with more intense stellar winds \citep{Wright2015a}.
Names are omitted for main sequence O- and B-type stars that have less massive winds. 
The X-ray mosaic intensity is logarithmically scaled.} 
\label{hr_mosaic}
\end{figure*}

According to simple theoretical scaling approximations \citep{Stevens2003}, 
part of the stellar wind kinetic energy is thermalized, so we can derive a simple estimate 
for the expected temperature of the shocked gas (T$_{\rm shock}$) 
$\approx$ 1.3$\times$10$^{7}$\,($v_{\rm w}$/1000)$^2$ in K, 
where $v_{\rm w}$ is the typical stellar wind velocity of massive stars in units of km\,s$^{-1}$. 
For massive stars the stellar wind expansion obeys a $\beta$-law velocity, 
$v(r)=v_\infty(1-R_\star/r)^\beta$, with $\beta$=0.8 for super-sonic 
winds \citep{Pauldrach1986}, thus for distances greater than 10 times the 
stellar radius (10R$_\star$ $\le$ 0.01 pc) the ratio 
$v_{\rm w}/v_\infty \ge 0.95$, so $v_{\rm w}$ is well approximated with typical $v_\infty$ values. 
The stellar terminal wind velocity of early O-type stars ranges from 900 to 2800~km~s$^{-1}$ so 
gas temperatures would range between 1.0 and 10.1 $\times 10^{7}$~K, which is equivalent to energies 
of $\sim$ 0.9 to 8.7 keV.  While these energies are roughly consistent with what we observe,
they should be considered as upper-limits because T$_{\rm shock}$ here is computed as a purely 
radiative limit. In a more realistic description, it is expected that a fraction of 
 energy released  from the winds of massive stars 
 deposits mechanical  energy into the ISM. 

The entire massive stellar content of the Cygnus OB2 region comprises: $i)$ 25 evolved 
(of Classes I, Ie, II((f)), III) massive stars of O-type, plus 4 Wolf-Rayet stars; $ii)$ 28 evolved B-type 
stars of classes I, Ia, Ib, II, III, most of them characterized by slow winds and faint, or absent, 
intrinsic X-ray emission; $iii)$ 113 O-type and early B-type stars on the main sequence  \citep{Wright2015b}.
The contribution to the total stellar wind energy L$_{\rm w}$ released by each star 
can be obtained according to the expression 3$\times$10$^{35}\,\dot{M}_{\rm -6}\,$($v_{\rm w}/1000)^2$ in erg\,s$^{-1}$, 
where $\dot{M}_{\rm -6}$ is the mass loss rate in units of $10^{-6} M_\odot$~yr$^{-1}$, \citep{Canto2000}. 
We adopted individual mass loss rates according to  the \cite{Vink2001} formalism 
and terminal wind velocities assumption of v$_\infty$= 2.6\,v$_{\rm esc}$ \citep{Lamers1995}, which 
were computed by \cite{Rauw2015} for the entire massive star population of the region. 
The total L$_{\rm w}$ released by each one of these three groups separately is 
$\approx$\,1.3$\times$10$^{38}$,  6.6$\times$10$^{36}$, and 1.3$\times$10$^{37}$ erg\,s$^{-1}$, respectively. 
The total stellar wind energy injected into the ISM of the region, from 
the entire population of massive stars, is then $\approx$\,1.5$\times$10$^{38}$ erg\,s$^{-1}$, 
with the evolved O stars dominating.

At this point, it is interesting to estimate the kinetic to diffuse X-ray emission efficiency by computing the
diffuse X-ray luminosity of the region.
Early works based on the CWM have provided some reliable estimates for the X-ray diffuse emission in massive
SFRs, e.g. R136 and NGC\,3603 \citep{Moffat2002}, NGC\,346 \citep{Naze2002}, 
Rosette \citep{Townsley2003}, and Westerlund I \citep{Muno2006}. They have found that the diffuse X-ray 
luminosity in the broad band of 0.5--8.0 keV usually lies in the range (1-6) $\times$10$^{34}$ erg\,s$^{-1}$. 
Such estimates indeed confirm the efficiency L$_{\rm x}^{\rm diff}$\,/\,L$_{\rm w}$ $\sim$
2$\times$10$^{-4}$, as predicted by \cite{Dorland1987} through the investigation of dissipative mechanisms.  
Now, if we assume that the diffuse X-ray emission luminosity is entirely produced  through wind shock--ISM dynamical 
interaction, the efficiency in Cygnus~OB2 is $\eta$ $\sim$ L$_{\rm x}^{\rm diffuse}$ / L$_{\rm w}$ = 
4.2$\times$10$^{34}$ / 1.5$\times$10$^{38}$ = 2.8$\times$10$^{-4}$, or about a factor $\sim$\,2  
more efficient than in Westerlund 1 (Wd\,1).

Next, we are able to estimate the diffuse X-ray luminosity via
the CWM model.  Such a model assumes that the bulk of the diffuse emission in the region is  
due to a hot plasma that exhibits relaxed, center-filled morphology with a lack of any obvious, 
measurable temperature gradients. We therefore considered the simple hypothesis of uniform,
optically thin thermal plasma with a simple geometry, although the emission doubtless 
has a more complex structure in reality.
As discussed in section\,\ref{hrsection}, thermal diffuse emission is an adequate description 
if we adopt diffuse emission regions with HR below -\,0.1. Figure\,\ref{hr_mosaic} shows this 
diffuse emission with regions of harder emission removed.

The total mass of hot gas (HR$\ge$-0.1) in the region that emits in X-rays covers  overly precise $\sim$35 \% 
of the total survey area of $\approx$ 0.97 sq-deg, which at the distance of 1450 pc to Cygnus OB2, 
corresponds to an area of 209.5 pc$^2$. This area is equal to a 14.5$\times$14.5 parsec side (s$_x$) square. 
Because stellar wind--ISM interaction occurs in a three-dimensional (3D) space, 
the observed bi-dimensional (2D) diffuse gas density in (cm$^{-2}$) cannot directly be compared with
gas density in cm$^{-3}$ units. We therefore converted the observed 2D to 3D geometry, by 
conservation of the total mass$--$square area equal to sphere surface$--$of a 2D circle 
to a 3D sphere of the diffuse gas that radiates in X-rays. 
Thus the area s$_x^2$=4.$\pi$\,R$_{\rm c}^2$, where R$_{\rm c}$ is the cluster effective radii. This equation gives 
a R$_{\rm c}$ of $\sim$\,4.1 pc ($\sim$ 0.16 deg), so the characteristic plasma volume of the region is 
V$_{\rm x}$=4/3$\pi$R$_{\rm c}^3$ = 8.48$\times$10$^{57}$ cm$^{-3}$.
Using the CWM and the analytic solutions to the density (n$_{\rm 0}$) in cm$^{-3}$ 
(see equation 4 of \cite{Canto2000}), we find n$_{\rm 0}$=0.06 cm$^{-3}$ that is
the typical mass density contribution from massive stelar winds in the region. 

The emission measure of the diffuse X-ray emission is obtained by integration 
of electron density squared over the emitting volume (EM = 3/4$\pi$R$_{\rm c}^3$\,n$_{\rm 0}^2$), 
thus EM\,$\sim$\,1.71$\times$10$^{55}$ cm$^{-3}$, and XSPEC normalization (Norm =
10$^{-14}$\,EM/(4$\pi\,D)^2$) of 3.9. With all these parameters, 
and assuming that the emission is well described by a combination of 
three thermal plasmas (see section\,\ref{spectral}),
we simulated a fake X-ray spectrum by assuming APEC models at temperatures
kT$\approx$ 0.1, 0.4 and 1.2 keV (see table\,\ref{spectral_tab}).
In order to compute theoretical absorption-corrected X-ray luminosity, 
we applied individual multiplicative neutral Hydrogen absorption column (using TBabs) to the emission models, 
Values of N$_{\rm H}$=0.42, 1.1, and 1.3 $\times$10$^{22}$ cm$^{-2}$ were obtained 
from our X-ray spectral fitting results (see sub-section\,\ref{spectral}). 
It should be mentioned that if we adopt the dust column density relationship 
N$_{\rm H}$/A$_{\rm v}$=1.6$\times$10$^{21}$ cm$^{-2}$, which represents the better proxy 
for the soft X-ray absorption column density from HI maps (Flaccomio et al., this issue),
the A$_{\rm v}$ in the region ranges between 2.6 and 6.8 mag, values which are consistent with the median 
of value 4.5 mag (Guarcello et al. this issue).   
Using these approaches, we predict a theoretical soft X-ray luminosity 
L$_{\rm x}^{\rm Soft}$=2.1$\times$10$^{\rm 34}$ erg\,s$^{-1}$, and hard 
L$_{\rm x}^{\rm Hard}$=0.2$\times$10$^{\rm 34}$ erg\,s$^{-1}$, leading to a 
total diffuse X-ray luminosity L$_{\rm x}^{\rm diff}$ of 2.3$\times$10$^{34}$ erg\,s$^{-1}$, 
for the 0.5-7.0 keV energy band, a factor $\sim$2 lower than spectral fit results.  
In any case, and for different reasons, diffuse X-ray luminosity computed from spectral fitting or
from CWM, should be more rigorously considered a lower limit.

Comparing our X-ray flux estimate to that of other SFRs, we find the diffuse X-ray luminosity in Cygnus OB2 
is a factor $\sim$3 larger than that estimated for the Arches cluster (1.6$\times$10$^{\rm 34}$ erg\,s$^{-1}$, 
\cite{Yusef2002}); $\sim$2 times larger than for the massive NGC 3603 cluster (2.0$\times$10$^{\rm 34}$ erg\,s$^{-1}$, 
\cite{Moffat2002}); but $\sim$7 times fainter than computed for the Carina Nebula 
(3.2$\times$10$^{\rm 35}$ erg\,s$^{-1}$, \cite{Townsley2011a}) for the same energy band.
These agreements support the idea that the massive stellar content plays a major, but proportional, 
role in the efficiency of conversion from injected kinetic wind energy through 
ISM interaction to the diffuse X-ray luminosity, that for Cygnus OB2 is $\sim$\,3$\times$10$^{-4}$.

However, it is plausible that non-thermal processes may be acting 
efficiently near the massive stars, altering, and perhaps increasing, the diffuse X-ray 
luminosity of the region. Thus, it is of interest to examine potential non-thermal 
emission contributions and their implications in a multi-wavelength context.

\subsection{Non-thermal contribution to diffuse X-ray emission}

Diffuse X-ray emission may also be produced through non-thermal mechanisms. 
Evidence for non-thermal processes has been uncovered in the Westerlund~1 star cluster 
by \citet{Muno2006}. However, those authors suggested that about 30\% of the diffuse  X-ray emission 
continuum would be produced by the unresolved Pre Main Sequence star (PMSs) population in the region, 
probably through magnetic reconnection flares and/or micro-flares. The relevant 
non-thermal emission processes are Synchrotron Losses (SL) and 
Inverse Compton (IC) scattering, which naturally produce more hard ($\ge$ 2 keV) X-rays, and are 
consequently not trivially distinguishable from the AGN background diffuse contribution. A third process, 
leading to softer X-rays, is Charge Exchange (CXE) line emission. The relative importance 
of these mechanisms depends on how, and where, the required population of non-thermal particles 
are created in the region, the neutral Hydrogen density of the local ISM, and also the local magnetic field 
in the region.

\begin{figure}[!h]
\centering
\includegraphics[width=8cm,angle=0]{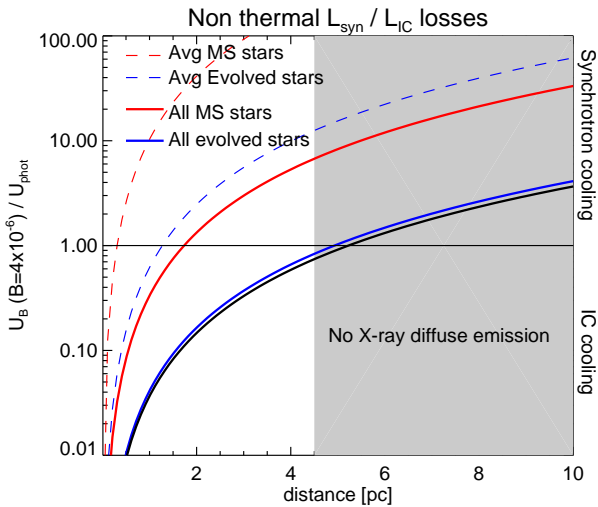}
\caption{\small Curves representing the expected relative contribution of synchrotron to inverse Compton loses to the diffuse
X-ray emission. The black continuous line corresponds to the ratio U$_{\rm B=4 \mu\,G}$ / U$_{\rm ph}$ accounting for the 
total contribution from all massive stars of the region with respect to the centre of the diffuse emission.  The coloured grey
region represents distances from the adopted centre (RA= 20:33:00 ,  DEC=41:20:00) 
in which diffuse X-ray emission is not observed.}
\label{losses}
\end{figure}

In order to compare relative SL and IC loses, we compute the ratio between the radiation field density (U$_{\rm ph}$) 
of massive stars of the region  and the expected ISM magnetic density (U$_{\rm B}$=B$^2$/8$\pi$) of the region.
We adopted a typical ISM magnetic field of B$\sim$4 $\mu$G \citep{Beck2001} to calculate U$_{\rm B}$
and used the individual bolometric luminosity (L$_{7}$) to compute U$_{\rm ph}$ 
(= 5.5$\times$10$^{-9}$L$_{7}$d$_{\rm pc}^{-2}$) in erg\,cm$^{-3}$ around each massive star \citep{Muno2006}. 
The individual U$_{\rm B}$/U$_{\rm ph}$ ratio as a function of distance (d$_{\rm pc}$) for each 
massive star was calculated, but for simplicity Figure\,\ref{losses} only shows the median and 
integrated U$_{\rm ph}$ as a function of distance from the stars. The averaged 
U$_{\rm B=4 \mu\,G}$ $\ge$ U$_{\rm ph}$ condition is supplied for distances above 0.5 pc
and 1 pc for single main sequence (MS) and evolved massive stars, respectively.
We considered the projected density of massive stars in the region, which has a typical 
value of 0.8 star/pc$^2$, so
the respective contributions to U$_{\rm ph}$ from individual massive stars should be added 
to get a more realistic estimation of the U$_{\rm ph}$. The radiation field hypothetically 
has a maximum at the centre of the diffuse X-ray emission --in agreement with the spatial 
density of massive stars-- (centred at RA= 20:33:00 ,  DEC=41:20:00), see Figure\,\ref{hr_mosaic}.
Figure\,\ref{losses} shows that SL becomes important only 
for distances larger than $\sim$ 5 pc ($\sim$ 12.4 arcmin) from the centre of the region. However, at such a 
distance, the observed diffuse X-ray emission is absent, or just marginally detected in soft X-rays, 
so that synchrotron losses would not be contributing significantly to the observed X-ray diffuse emission.

\label{NT}

Otherwise, Inverse Compton (IC) scattering is a potential loss term for diffuse X-ray emission, 
feeding off copious UV photons from massive stars. However, the dilution of the UV field (U$_{\rm phot}$) 
increases with the distance to the stellar source,  and thus rapidly decreases the energy of the electrons
after they leave the vicinity of the shock, imposing a natural "short" distance restriction for the action of this mechanism.
So, considering that the luminous (evolved) stars shock occurs in the radiative cooling limit, we compared  
synchrotron  to IC loses through the expression 
L$_{\rm syn}$/L$_{\rm IC}$ = U$_{\rm B}$/U$_{\rm phot}$\,$\approx$\,7.1$\times$10$^{-4}$B$^2$/L$_{6}$, where
B=B$_{\rm star}$(v$_{\rm rot}$.R/v$_\infty$)d$^{-1}$  
and L$_{6}$ in 10$^6$L$_\odot$ \citep{White1995}. 
As U$_{\rm syn}$ and U$_{\rm phot}$  go as the inverse square of the distance, and 
adopting typical values for the magnetic fields of the massive stars (B$_{\rm star}$ $\sim$ 200 G), 
L$_{\rm syn}$/L$_{\rm IC}$\,$\approx$10$^{-7}$, so the
IC process is dominant over SL near the massive stars.
However, the energy requirement to raise the input energy of UV photons
(E$_{\rm input}$ $\sim$ 10 to 20 eV) into the observed X-ray regime of the diffuse emission (E$_{\rm  out}$ 
in 0.5 - 7.0 keV) requires a population of accelerated electrons of moderate energy ($\sim$ 5-35 MeV)
generated in the inner wind of massive stars, or even from colliding wind regions \citep{Muno2006}. 
So IC cooling may  act in the inner stellar winds and cannot travel far from the 
acceleration site \citep{Chen1992, Eichler1993}; thus, large-scale (few parsecs) IC 
cooling is not expected to contribute to the observed diffuse X-ray emission. 

These theoretical predictions for insignificant contributions from synchrotron losses and 
inverse Compton scattering therefore agree with the observed absence of hard [2.5\,-\,7.0] 
keV diffuse X-ray emission (Figure\,\ref{largescale}).

\begin{figure*}[!ht]
\centering
\includegraphics[width=18cm,angle=0]{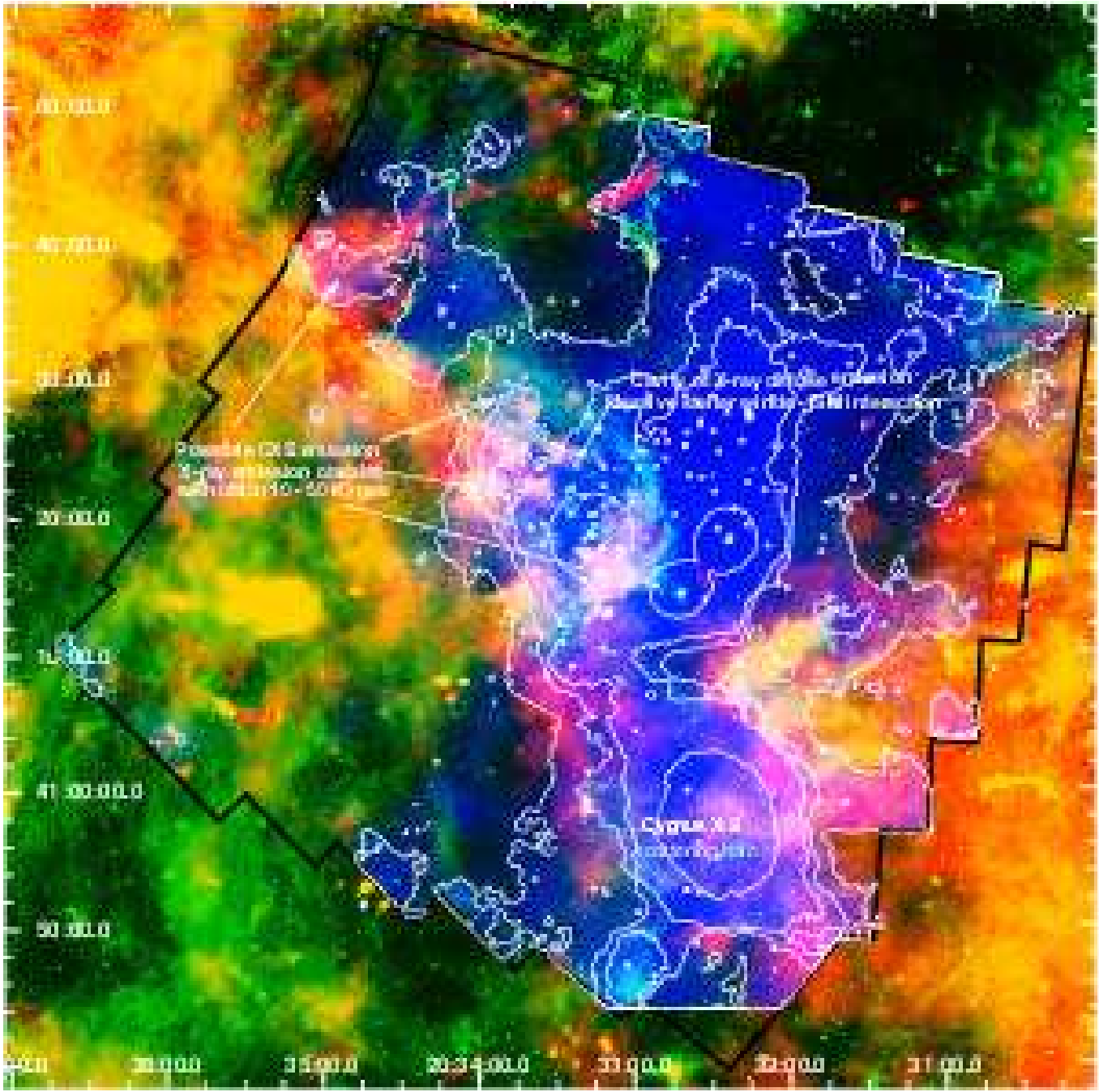}
\caption{\small The neighbourhood of 
Cygnus OB2. The ACIS-I mosaic of the Cygnus OB2 survey is outlined in black. 
Small circles indicate the massive OB star content of the region, regardless of spectral type 
and evolutionary class \citep{Wright2015a}. Diffuse X-ray contours corresponding to flux levels 
of 3.0, 3.7, 4.1, 4.7 and 8.3 $\times$10$^{-10}$ ph/cm$^2$/s/arcsec$^2$, are shown in white.
The false RGB color image was composed as follows: 
The Herschel 500\,$\mu$m (T\,$\approx$\,10 K) cold gas emission in red \citep{Schneider2016}, 
 the 8\,$\mu$m Spitzer IRAC image for the warm gas (T $\approx$150 K) in green, and the diffuse X-ray emission in the
0.5\,--\,2.5 keV energy range in blue. 
There are also four arrows that indicate the regions which show signs of the coexistence of 10 K cold 
ISM gas and hot diffuse X-rays, which are likely scenarios for CXE emission.
}
\label{whr}
\end{figure*}

Alternatively, there is one further mechanism that could produce non-thermal diffuse X-ray emission. 
Theoretical \citep{Wise1989}, and more recently observational 
\citep{Townsley2011a}, considerations suggest highly charged ions associated with hot 
plasma (from massive stellar winds) could interact with neutral atoms of ambient cool 
or warm ISM gas via the Charge X-ray Exchange (CXE) mechanism.
CXE is a non-thermal line emission mechanism that could produce conspicuous 
diffuse X-ray emission at soft (line-emission) energies, even when such interactions occur 
on spatial scales of several tenths of parsecs \citep{Montmerle2012}. Calculations from 
\cite{Wise1989} have shown that CXE emission is negligible in the case of Supernova Remnants 
with fast shocks, but becomes important for less energetic cases, such as hot gas interacting 
with cold, dense ISM structures. The process is more efficient for lighter elements \citep{Lallement2004} and 
a wide variety of emission lines below 2 keV are expected from elements in various ionization states, 
with no continuum contribution. In Carina, for example, lines and possible elements responsible for them are 
0.64 keV (O), 0.77 keV (O or Fe), 1.07 keV (Ne), 1.34 + 1.54 keV (Mg), 1.80 keV (Si), and some faint
lines at harder X-ray energies such as 2.61 keV (S), and 6.50 keV (Fe) \citep{Townsley2011a}. 
Such emission has been also suggested in other massive star-forming regions for which the data were
of sufficient quality \citep{Townsley2011b}.
In Figure\,\ref{whr} we inspected the neighbourhood of 
Cygnus OB2 in a multi-wavelength approach, by searching for plausible regions for CXE emission, where 
soft X-ray diffuse emission co-exists with "warm" ($\approx$ 100-150 K) and cold ($\approx$ 10-50 K) 
gas structures observed in the infrared  ({\it Spitzer} and Herschel data, respectively). 
Just four zones satisfy this condition, one to the north and the other toward the center of the region.
However, soft diffuse emission luminosity in these three regions is low enough to search for narrow band 
images at excesses produced by CXE emission line energies, e.g. $\sim$[0.6-0.8] keV for O and/or 
Fe lines; $\sim$[1.0-1.6] keV for Ne and/or Mg. 
Although the poor photon statistics in the X-ray spectra of these regions do not allow us confirm
the presence of line emission from He-like and H-like states of the elements C, N, and O, the mere existence
of soft diffuse X-ray emission, even far from the massive stars, would be considered a favorable place for 
the CXE mechanism occurring in Cygnus OB2.

The multi-wavelength image of the region (see Figure\,\ref{whr}) also shows 
that most X-ray diffuse emission is spatially coincident with regions of low infrared surface brightness, 
which is in coincidence with regions of low ISM extinction. This suggests the X-ray plasma is  
volume-filling, rather than edge-brightened, just as was found for the Carina Nebula  \citep{Townsley2011a} 
and other giant H II regions \citep{Townsley2011b}.
This is an indirect observational probe that in the case of Cygnus OB2, powerful stellar winds 
from massive stars primarily  collide between the OB winds rather than independently with 
the exterior cold cloud, which is in concordance with \cite{Canto2000} 3D modelling of SFRs with
high density population of massive stars.\\

Finally, and for the first time, we have resolved X-ray diffuse emission haloes at sub-parsec scales around 
some evolved massive stars of the region (see figures and respective comments in the Appendix). 
We defer a more detailed study of this phenomenon to future work.\\

\section{Summary}

A thorough and detailed analysis of 40 {\it Chandra} ACIS-I observations of the Cygnus OB2 Association, 
including the removal of the 7924 X-ray point-like sources detected, has revealed the diffuse X-ray 
emission that permeates the region. We have mapped a region $\sim$ 30 pc$^{2}$ across 
at a spatial resolution reaching down to a few thousand of AU. 
The main findings of the study of this diffuse emission are as follows.

\begin{enumerate}
\item Large-scale X-ray diffuse emission was seen in the broad 0.5\,--\,7.0 keV energy band, and was 
also detected in the Soft [0.5\,:\,1.2]  and Medium [1.2\,:\,2.5]~keV bands. 
A marginal detection of diffuse emission was made in the Hard [2.5\,:\,7.0] keV band. 

\item The total diffuse X-ray emission luminosity  was found to be 
L$_{\rm x}^{\rm diff}$\,$\approx$4.2$\times$10$^{\rm 34}$ erg\,s$^{-1}$
(0.5-7.0 keV), and was well-represented by a three-component thermal plasma 
model with typical temperatures of kT$\approx$ 0.11, 0.40 and 1.18 keV 
(1.2, 4.9 and 14 MK, respectively).

\item The extended moderate energy emission likely arises from O-type star winds thermalized 
by wind-wind collisions in the most populated regions of the association, while the Super-Soft (SS) emission 
probably arises from less energetic termination shocks against the surrounding 
interstellar medium (ISM).  The SS diffuse emission appears more dispersed than that at soft and medium energies, 
indicating diffusive motions of hot gas on 2 to 3 parsec scales.

\item The H{\sc i} absorption column density was constrained with three individual $N_H$ models,
that on stacked spectra are N$_{\rm H}$ = 0.42, 1.12 and 1.30 $\times$10$^{22}$ cm$^{-2}$.  
At the center of the region, where most of the 
massive stars are located, N$_{\rm H}$ seems to be slightly lower than found for outer regions.
The diffuse X-ray emission is then spatially coincident with low extinction regions (and low ISM densities), 
that we attribute to powerful stellar winds from massive 
stars and their interaction with the local ISM. 
It is volume-filling, rather than edge-brightened, as has been found for other star forming regions.

\item An assessment of potential non-thermal diffuse emission sources finds that both synchrotron 
and inverse Compton scattering are not likely to contribute significantly to the observed large 
scale diffuse emission.  
A full assessment of a possible charge-exchange emission signal is challenging due to the 
large extinction that renders soft X-ray emission difficult to observe. By the way, this would require a more detailed spectral analysis.

\item Examination of the diffuse emission maps on smaller scales reveals  X-ray halos 
around evolved massive stars. This is the first time such emission structures have been detected.
\end{enumerate}

The results presented here highlight the value of large-scale X-ray surveys for 
understanding the energetics and feedback in massive star-forming regions, in addition to 
assessing their otherwise hidden or inconspicuous stellar content.\\

\acknowledgments

We thank the referee Dr. David Helfand for many corrections and very helpfully suggestions to our article.
JFAC is a researcher of CONICET and acknowledges their support. The research leading to these results 
has received funding from the European Union Horizon 2020 Programme under the AHEAD project (grant agreement n. 654215).
JJD and VK were supported by NASA contract NAS8-03060 to the {\it Chandra X-ray Center} (CXC) 
and thank the director, B. Wilkes, and the science team for continuing support and advice. 
MGG and NJW were supported by Chandra grant GO0-11040X during the course of this work. 
MGG also acknowledge the grant PRIN-INAF 2012 (P.I. E. Flaccomio).
NJW acknowledges a Royal Astronomical Society research fellowship and an STFC 
Ernest Rutherford Fellowship (grant number ST/M005569/1).
N.S. acknowledges support by the french ANR and the german DFG through the project "GENESIS" (ANR-16-CE92-0035-01/DFG1591/2-1).
We kindly thanks Dr. Patrick Broos of the Penn State UniversityÕs astrophysics group who share some private AE scripts for the 
diffuse X-ray analysis of the Chandra ACIS-I observations.
This research made use of Montage, funded by the National Aeronautics and Space Administration's
Earth Science Technology Office, Computation Technologies Project, under Cooperative Agreement Number
NCC5-626 between NASA and the California Institute of Technology.
Montage is maintained by the NASA/IPAC Infrared Science Archive.\\

\bibliographystyle{aa}
\bibliography{yaReferences}

\appendix
Diffuse X-ray emission of individual ObsID maps:\\

 In Figures 18 to 21, we show the individual diffuse X-ray maps (flux and hardness ratio) 
for the entire set of observations performed for the Cygnus OB2 Chandra ACIS-I survey. Each map is 17 sq.arcmin. 
The first and third columns show 
flux in the 0.5-7.0 keV energy band. Using the thermal model of the X-ray spectral fit (see section\,\ref{spectral}), 
we have computed the validated photon keV-to-erg conversion in the [0.5-7.0] keV band to be 
1 ph\,=\,4.18$\times$10$^{-9}$ ergs, which allows for conversion of diffuse maps of observed photon counts 
to absorbed flux. The X-ray flux is in CGS photon units.
Hardness ratio images are displayed in the second and fourth columns, respectively. 
They were computed as the ratio (S-H)/(S+H). True diffuse emission was only 
considered whenever $HR > -0.1$, corresponding to the colors green, yellow and red in the $HR$ maps
(see section 4 for details).
Intensities are illustrated on a single color scale of log(L$_{x}$)=[-17.08,-16.63] and HR=[-0.3,0.5] 
to facilitate comparison between flux and HR maps, respectively.\\
The filled and open symbols all over the maps refer to massive stars with and without intrinsic
X-ray emission, respectively. Stars correspond to evolved objects, while circles represent MS stars.
X symbols refer to massive OB type stellar candidates from \cite{Wright2015b}.

NOTE: DUE TO LARGE NUMBER OF FIGURES IN THE APPENDIX AND DISK USE LIMITS IN ASTRO-PH, MAP WILL BE 
ONLY AVAILABLE ON THE ApJSS (CYGNUS OB2 SPECIAL ISSUE) VOLUME.
\end{document}